\newif\ifconfver
\confverfalse      %declaring conference version false
\confvertrue        %declaring conference version true

\ifconfver
     \documentclass[10pt,twocolumn,twoside]{IEEEtran}
\else
    \documentclass[11pt,draftcls,onecolumn]{IEEEtran}
\fi

\usepackage{graphicx,booktabs,color}  % Written by David Carlisle and Sebastian Rahtz
\usepackage{amsmath,epsfig,amsfonts,amssymb,graphics,psfrag,amsthm,calc,subfigure,url,bm,cite}
\usepackage{algorithm,algpseudocode}
\usepackage{float}
\usepackage{stfloats}  % Written by Sigitas Tolusis
\usepackage{cases,epstopdf}
\newlength{\twidth}
\ifconfver
\else
\fi

    \makeatletter
    \def\multilimits@{\bgroup
  \Let@
  \restore@math@cr
  \default@tag
 \baselineskip\fontdimen10 \scriptfont\tw@
 \advance\baselineskip\fontdimen12 \scriptfont\tw@
 \lineskip\thr@@\fontdimen8 \scriptfont\thr@@
 \lineskiplimit\lineskip
 \vbox\bgroup\ialign\bgroup\hfil$\m@th\scriptstyle{##}$\hfil\crcr}
    \def\Sb{_\multilimits@}
    \def\endSb{\crcr\egroup\egroup\egroup}
\makeatother

\definecolor{orange}{RGB}{255,107,0}

\newtheorem{Lemma}{Lemma}

\newtheorem{Theorem}{Theorem}

\newtheorem{Fact}{Fact}
\theoremstyle{remark}

{\begin{list}{}%
    {\setlength{\rightmargin}{0pc}%
    \setlength{\leftmargin}{1pc} }} %
{\end{list}}

{\begin{list}{}%
    {\setlength{\rightmargin}{1pc}%
    \setlength{\labelsep}{0pc}
    \setlength{\leftmargin}{1pc} }} %
{\end{list}}

{\begin{list}{}%
    {\setlength{\rightmargin}{0pc}%
    \setlength{\leftmargin}{0pc} }} %
{\end{list}}

{\begin{list}{}{
    \settowidth{\labelwidth}{\mbox{\textnormal{#1}}}%
    \setlength{\leftmargin}{\labelwidth+\labelsep}}}%
{\end{list}}

\begin{document}

\bibliographystyle{IEEEtran}

\title{Latency Minimization  for Multiuser Computation Offloading  in  Fog-Radio Access Networks}
	%Constant Modulus Multicast Beamforming for Multiuser Massive MISOME  Wiretap Channel}

\ifconfver \else {\linespread{1.1} \rm \fi

\author{
Wei Zhang, Shafei Wang, Ye Pan, Qiang Li, Jingran Lin and Xiaoxiao Wu
\thanks{	
W. Zhang, S. Wang, Y. Pan, Q. Li and J. Lin are with the School of  Information and Communication Engineering, University of Electronic Science and Technology of China, P.~R.~China,
Chengdu, 611731.

S. X. Wu is with the College of Information Engineering, Shenzhen University,
Shenzhen 518060, China.}}

%\vspace*{-2\baselineskip}

\maketitle

\ifconfver \else
\begin{center} \vspace*{-2\baselineskip}
%11th Revision, \today \\[2\baselineskip]
\end{center}
\fi
\begin{abstract}
This paper considers computation offloading in fog-radio
access networks (F-RAN), where multiple user equipments
(UEs) offload their computation tasks to the F-RAN through a
number of fog nodes. Each UE can choose one of the fog nodes
to offload its task, and each fog node may serve
multiple UEs. Depending on the computation burden at the fog
nodes, the tasks may be computed by the fog nodes or further
offloaded to the cloud via capacity-limited fronthaul links. To
compute all UEs’ tasks as fast as possible, joint optimization
of UE-Fog association, radio and computation resources of
F-RAN is proposed to minimize the maximum latency of all
UEs. This min-max problem is formulated as a mixed integer
nonlinear program (MINP). We first show that the MINP can
be reformulated as a continuous optimization problem, and
then employ the majorization minimization (MM) approach to
find a solution. The MM approach that we develop is unconventional in that---each MM subproblem can be solved {\it inexactly} 
with the same provable  convergence
guarantee as the conventional exact MM, thereby reducing the complexity of each MM iteration. In addition, we
also consider a cooperative offloading model, where the fog
nodes compress-and-forward their received signals to the
cloud. Under this model, a similar min-max latency optimization problem is formulated and tackled again by the inexact
MM approach. Simulation results show that the proposed
algorithms outperform some heuristic offloading strategies, and
that the cooperative offloading can better exploit the transmission diversity to attain better latency performance than the
non-cooperative one.
\\\\
\ifconfver
\else
\noindent {\bfseries EDICS}: MSP-CODR (MIMO precoder/decoder design), MSP-APPL (Applications of MIMO communications and signal processing), SAM-BEAM (Applications of sensor and array multichannel processing)
\fi
\end{abstract}

\ifconfver \else \IEEEpeerreviewmaketitle} \fi

%---------------------------------------------------------------------------
\ifconfver \else
\newpage
\fi

\section{Introduction} \label{sec:introduction}
The next generation wireless communication system is expected to provide ubiquitous connections for massive heterogenous Internet of Things (IoT) devices with high speed and low latency. The current cloud-computing-based network infrastructure is facing challenges to meet these requirements, because massive heterogenous requests with different data size and latency requirements need to be forwarded to and processed at the central baseband processing units (BBUs), which, however, could cause heavy burden on the fronthaul, and incur intolerable latency for some delay-critical missions. For example, in
some interactive applications, e.g., virtual reality, industrial automation and  vehicle-to-vehicle communications,
the round-trip delay may be required below a few tens of milliseconds~\cite{Chiang}. To meet the critical latency requirement and alleviate the pressure on the fronthaul, a fog-computing-based radio access network (F-RAN) has recently been proposed as a promising solution~\cite{Peng}. The concept of F-RAN is developed from the fog computing, which was originally proposed by Cisco~\cite{Bonomi}. By shifting certain amount of computing, storage and
networking functions from the cloud to the edge of the network, F-RAN is able to provide  more prompt responses to users' requests with less fronthaul bandwidth occupation.

Evolving from cloud RAN (C-RAN) to F-RAN, the wireless access point (AP) is endowed with more capabilities and functions, such as computation and content caching. In this work, we focus on the computation aspect of F-RAN,  and investigate how the enhanced APs (also called fog nodes in the rest of the paper) near the user equipments (UEs)  can help improve the latency performance in the fog-assisted computation offloading applications. 
%Herein, the fog-assisted computation offloading means the computation-intensive tasks at the IoT devices are  offloaded  and computed either at the fog nodes or the cloud, depending on the computation loads at the fog nodes and the  capacities of the fronthaul links. 
Conventionally, computation offloading has been extensively studied in the context of mobile-edge computation (MEC)~\cite{Mach,Maray,NAYAK2022,SADATDIYNOV2022}. MEC considers that there is one or multiple computing servers to process the tasks, which are partially or wholly offloaded by UEs. The offloading is usually accomplished via wireless transmissions from UEs to the MEC server, and the UEs are competing with each other for the radio and computation resources.  To provide  satisfactory quality-of-service (QoS) for UEs, a joint optimization of the offloading decision and resource allocation is the  crux of achieving efficient MEC.

Earlier studies on MEC focused on the offloading decision-making for single UE admission. By assuming infinite computation capacity of the  server, the trade-off between the offloading and local computation was thoroughly investigated in~\cite{Odessa,WWZHANG13,WWZHANG15}. Recently, more efforts have been devoted to joint optimization of offloading decision, communication  and computation resource allocations. Typically, this kind of problems are formulated as mixed-integer nonlinear programs (MINP) with different utility functions. In~\cite{Cai_vtc,Lyu,FWang18} the authors studied the MEC problem with the goal of minimizing the total energy consumption, including transmission and computation energy, subject to UEs' latency requirements. CCCP~\cite{Cai_vtc}, quantized dynamic programming~\cite{Lyu} and Lagrangian duality method~\cite{FWang18} are employed to find approximate solutions for the MINP. In~\cite{qliu_GC18,Ren18,Du_tvt19,Saleem20,KUANG21,LI2022247}, the latency is adopted as the system utility function. The work~\cite{qliu_GC18} developed an iRAR algorithm to minimize the sum latency for multiple base stations (BSs) and multiple computing servers. For the case of single computing server, the work~\cite{Ren18}  derived the optimal resource allocations under local computing, cloud computing and mixed computing models. Different from~\cite{qliu_GC18,Ren18}, the work~\cite{Du_tvt19,LI2022247} studied the worst-case latency minimization problem in order to provide latency fairness for UEs. By extending the fireworks algorithm to the binary case,  a heuristic offloading decision and resource allocation scheme was proposed. To  balance energy consumption and latency, the weighted energy-plus-latency utility function is also commonly adopted in  MEC offloading~\cite{Tran,Du18,Dong_tbc18,Leung18,LIU2022}. 
%By setting different weighting coefficients for energy consumption and latency, this model has flexibility to balance latency-critical and energy-efficient offloading. 
Apart from the above models, there are also other MEC models, which are proposed to address some specific issues in offloading, such as dynamic environment change, online and distributed implementations of offloading schemes;  see~\cite{QLing18,Lin1,Lin2,Lin3,A1,A2,A3} and the references therein.

\begin{figure}[!h]
	\centerline{\resizebox{.4\textwidth}{!}{\includegraphics{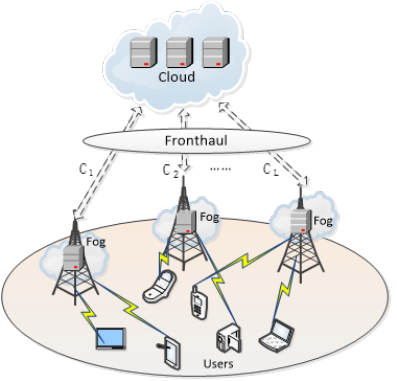}}
	}  \caption{F-RAN model.} \label{fig:model}
	%\vspace*{-1.4\baselineskip}
\end{figure}

Back to F-RAN, this work focuses on the fog-assisted computation offloading. Different from the above MEC models, the  fog-assisted computation offloading model consists of three layers, the UE layer, the fog layer and the cloud layer; see Figure~\ref{fig:model} for an illustration. Each UE offloads its computation task to F-RAN via one of the fog nodes. The tasks may be processed by the fog nodes or further offloaded to the cloud, depending on the computation and the fronthaul capacities of the fog nodes. To guarantee fairness,  a min-max latency minimization criterion is adopted herein to  optimize the F-RAN resources---which include the UE-Fog association, radio and computation resources---so that the worst latency of all UEs induced by transmission and computation is as small as possible. This min-max latency optimization problem is formulated as an MINP. With a careful treatment of the binary variables, we show that the MINP can be equivalently reformulated into a form involving only continuous variables, and thereby powerful machinery in continuous optimization can be exploited to handle it. Specifically, by incorporating the idea of majorization minimization (MM)~\cite{MM} and the weighted MMSE (WMMSE) reformulation~\cite{Shi}, we develop an {\it inexact} MM algorithm for the min-max problem with   convergence guarantee to a Karush-Kuhn-Tucker (KKT) solution.

We should mention that the aforementioned min-max fairness problem assumes that each fog node individually forwards the associated UE's task to the cloud, if the task is processed at the cloud. To fully capture the cooperative gain of the fog nodes, we also consider a cooperative offloading strategy, where all the fog nodes  compress-and-forward their received signals  to the cloud. With cooperative offloading,
the UE-to-cloud channel can be seen as a virtual multiple-access channel (V-MAC). By applying a similar discrete-to-continuous variable reformulation, an inexact MM algorithm is developed to find a solution. Simulation results demonstrate that the cooperative offloading can generally provide better latency performance as compared with the non-cooperative one.

\subsection{Related Works and Contributions}
There are some related works worth mentioning. The works \cite{Aldo} and \cite{Davidson,Mostafa22} considered a joint optimization of radio and computation resources for energy minimization with latency constraints in single-cell and multicell networks, respectively, where all the computation is done at the cloud with the UE-BS association  prefixed. In~\cite{Pang16,Pang17}, a cooperative computation model is considered, but their focus is more on choosing appropriate number of fog nodes for each task, given the communication resource constraints.  The  latency minimization, the energy minimization and  the energy-plus-delay minimization are respectively considered in~\cite{Du18,Dong_tbc18,pan20}, \cite{Chang21} and \cite{Du_tvt19,PengMG19} under the setting of multiple UEs, one computing AP (or fog node) and a cloud server. The work~\cite{Tony20} studied the optimal computation task scheduling problem in order to minimize the total latency. Since there is only one computing AP, no UE-AP association optimization is needed and moreover transmit beamforming is not considered in~\cite{Du18,Dong_tbc18,pan20,Chang21,Du_tvt19,PengMG19,Tony20}. The  work~\cite{Leung18} deals with a similar problem as~\cite{Du18,Dong_tbc18,LA20193} under the multi-fog setting, however, beamforming and cooperative offloading among fog nodes is again not considered.  The work~\cite{Ma2020} studied a hybrid communication and computation offloading  problem in F-RAN by jointly optimizing the UE-AP association and the bandwith allocation. A genetic convex optimization algorithm (GCOA) was proposed to divide the original  MINP into two convex optimization problems.  Apart from the optimization-based approach, learning-based offloading decision approach has also attracted much attention recently. In particular, the works~\cite{Zhao20,Rahman20,Lee21,Fan21,ZhangLL22} proposed deep reinforcement learning and  federated learning methods to centralized  or decentralized learn the offloading policy.
Besides computation offloading decision, there are other works investigating F-RAN from various perspectives, including   the offloading performance analysis~\cite{Xu19,Jofina20,REN2021},  energy efficiency optimization~\cite{Ly21,Liu22} and cache deployment~\cite{Jiang21,hindia22}, to name a few; see~\cite{WangZ22} and the references therein.

To summarize, compared with the existing works on F-RAN, the main contributions of this work are as follows.
\begin{enumerate}
	\item We consider a general compuation offloading model in F-RAN, which include  transmit beamforming between multiple multi-antenna UEs and multiple multi-antenna fog nodes,  UE-fog association, fog-cloud computation task distribution and cooperative computation offloading among fog nodes. To the best of our knowledge, this comprehensive  model has not been touched in the current literature.  
	
	\item Aiming at providing  latency fairness for all UEs, we formulate a min-max delay problem by jointly optimizing the UE-fog association, the fog-cloud task distribution, radio and computation resource allocation. This min-max problem is a MINLP problem in its original form. We show that it can be equivalently transformed into a pure continuous optimization problem. Upon the latter, an {\it inexact} MM-based block-coordinate descent (BCD) method is proposed, and its convergence to a  KKT point is also proved.

	\item We have conducted extensive numerical simulations to demonstrate the efficacy of the proposed offloading scheme. Especially, simulation results reveal that under some conditions, the proposed cooperative offloading via compression-and-forward among fog nodes attains superior performance over the non-cooperative one. To the best of our knowledge, this  is the first work that investigates advantage of cooperative  offloading via compression-and-forward in the F-RAN\footnote{We should mention that compression-and-forward offloading was previously considered in~\cite{Chang21}, but that work focused on a single fog node without cooperative offloading.}.
\end{enumerate}

%Finally, we should mention as of writing of this paper, we are not aware of any computation offloading work, which takes into account the compress-and-forward offloading strategy in the fronthaul transmission. 

%Computation may be offloaded to the AP or further offloaded to the cloud.  A semidefinite relaxation (SDR) method is proposed to decide the tasks distribution and computational resource allocation for each task.

\subsection{Organization and Notations}
This paper is organized as follows. The system model and problem statement are given in Section~\ref{sec:model}.
Section~\ref{sec:approach} develops an inexact MM approach to tackling the min-max latency optimization problem.
Section~\ref{sec:soft_mode} considers a cooperative fog-assisted offloading model and develops an iterative algorithm to optimize the resources.
Simulation results comparing the proposed designs are illustrated in Section~\ref{sec:sim}. Section~\ref{sec:conclude} concludes the paper.

Our notations are as follows.
$(\cdot)^T$ and $(\cdot)^H$ denote the transpose and conjugate transpose, respectively;  $\bm {I}$ denotes an identity matrix with an appropriate dimension;  $\mathbb{C}^N$  denotes the set of complex vectors of dimension $N$; $\bm {A}\succeq \mathbf{0}$ (respectively $\bm {A}\succ \mathbf{0}$) means that $\bm {A}$ is Hermitian positive semidefinite (respectively definite); ${\rm Tr}(\cdot)$ denotes a trace operation; ${\rm Diag}({\bm A}, {\bm B})$ represents a block diagonal matrix with the diagonal blocks $\bm  A$ and $\bm B$; $\mathcal{CN}(\bm a, \bm \Sigma)$ represents a complex Gaussian distribution with mean $\bm a$ and covariance matrix $\bm \Sigma$.

\section{System Model and Problem Statement} \label{sec:model}
Consider an F-RAN, consisting of $K$ multi-antenna UEs, $L$ fog nodes and a cloud server. Each
UE has a computation task, however, due to limited computation capacity, all the tasks have to be offloaded to the F-RAN via the fog nodes.
%For simplicity,
%we assume that each user is connected with one fog node, and every fog node may serve multiple users. The association between the fog nodes and the users is not prefixed and needs to
%be jointly optimized with other  (communication and computation) resources so that
%the worst communication-plus-computation delay among the users is as small as possible.
%However, due to limited communication and computation resources of the F-RAN, the users have to compete with each other.
%In order to guarantee fairness among the users, we propose to study the F-RAN resource allocation problem from a max-min fairness perspective; i.e., we want to design a resource allocation scheme so that the maximum processing delay of all the users is minimized.
%\subsection{Task Offloading Model}
Suppose the user $k$'s task ${\sf T}_k$ is described by a two-tuple of $(D_k, B_k)$ integers, where  $D_k$ denotes the number of flops needed for completing ${\sf T}_k$, and $B_k$
represents the number of bits needed for encoding ${\sf T}_k$. To offload the task to F-RAN, user $k$ has to send the $B_k$ bits to the
fog nodes through wireless transmission. For simplicity,
we assume that each user gets access to F-RAN  through one of the  fog nodes, while each fog node may simultaneously provide access for multiple users. The association between the fog nodes and the users is not prefixed and needs to
be jointly optimized with other  resources.  To highlight this, we introduce a binary variable $\alpha_{k,\ell} \in \{0,~1\}$ to  indicate the association. In particular, $$\alpha_{k,\ell} = \begin{cases}
1, & \mbox{if user $k$ is connected with fog $\ell$}, \\
0, & \mbox{otherwise,}
\end{cases}$$
and $\sum_{\ell =1}^L \alpha_{k,\ell} =1, ~\forall~k\in {\cal K} \triangleq \{1,\ldots, K\}$.

Now, the offloading process can be described in the following two stages:

{\it Stage 1:~Wireless Transmissions from Users to Fog Nodes.} For ease of exposition, let us assume that user $k$ is associated with fog node $\ell\in {\cal L}\triangleq \{1,\ldots,L\}$, i.e., $\alpha_{k,\ell} = 1$ and $\alpha_{k,\ell'} = 0, \forall \ell' \neq \ell$. Let 
$$\bm x_k(t) = \bm v_k s_k(t) \in \mathbb{C}^{N_k},\quad k\in \cal K$$ 
be the transmit signal of UE $k$, where $\bm v_{k} \in \mathbb{C}^{N_k}$ 
is the transmit beamformer with $N_k$ being  the number of transmit  antennas, and $s_k(t)\in \mathbb{C}$ is the encoded signal for task  ${\sf T}_k$. Then, the received signal at the fog
$\ell$ is given by
\[ \bm y_\ell(t) = \bm H_{k,\ell}^H \bm v_k s_k(t) + \textstyle \sum_{j\neq k}  \bm H_{j,\ell}^H \bm v_j s_j(t) + \bm n_\ell(t),\]
where $\bm H_{j,\ell} \in\mathbb{C}^{N_j \times M_\ell}$ is the channel between  UE $j$ and fog $\ell$ with $M_\ell$ being the number of antennas at  fog $\ell$, and $\bm n_\ell (t) \sim {\cal CN}(\bm 0, \sigma_\ell^2 \bm I)$ is additive white Gaussian noise. The communication
rate between UE $k$ and fog $\ell$ is given by
\begin{equation}\label{eq:R_df}
%\begin{aligned}
R_{k,\ell} \hspace{-2pt}  = W \log ( 1 +  \bm v_k^H \bm H_{k,\ell} (\sigma_\ell^2 \bm I + \sum_{j\neq k}  \bm H_{j,\ell}^H \bm v_j \bm v_j^H \bm H_{j,\ell} )^{-1} \bm H_{k,\ell}^H \bm v_k )
%\end{aligned}
\end{equation}
where $W$ (Hz) is the bandwidth of the wireless transmission. The corresponding wireless transmission latency is
\begin{equation}\label{eq:delay_T}
\tau_{k,\ell}^T  = \frac{B_k}{R_{k,\ell}}.
\end{equation}
%Since user $k$ is served by only one fog node, we introduce a binary variable $\alpha_{k,\ell} \in \{0,~1\}$ to reflect the association relationship. In particular, $$\alpha_{k,\ell} = \begin{cases}
%  1, & \mbox{if user $k$ is served by fog node $\ell$}, \\
%  0, & \mbox{otherwise.}
%\end{cases}$$

{\it Stage 2:~Computing at the Fog Nodes/Cloud.} After the reception, the fog node $\ell$ may compute the task by itself or further offload the task to the cloud, depending on the fog's  computation load and the complexity of ${\sf T}_k$.
There are two cases:
\begin{enumerate}
	\item[1)] Computing ${\sf T}_k$ at the fog node. Let $f_{k,\ell}^F$  be the number of CPU flops allocated for executing ${\sf T}_k$ in every second. Then, the computation latency  is 
	\begin{equation}
	\tau_{k,\ell}^F = \frac{D_k }{f_{k,\ell}^F}.
	\end{equation}
	%where $\nu_{\ell}$ is the number of CPU cycles required for each floating point operation at fog $\ell$.
	
	\item[2)] Computing ${\sf T}_k$ at the cloud. In such a case, the processing latency consists of two parts. One is  transmission latency from the fog node to the cloud, and the
	other is the computation latency at the cloud. We consider that  fog $\ell$ is connected with the cloud via fronthaul with limited capacity $C_{\ell, \max}$ (bits/second). Let
	$C_{k,\ell}(\leq C_{\ell, \max})$ be the fronthaul capacity allocated by fog $\ell$ for further offloading ${\sf T}_k$ to the cloud. Then, the processing latency at the cloud is given by
	\begin{equation}\label{eq:tau_C}
	\tau_{k,\ell}^C = \frac{B_k}{C_{k,\ell}} + \frac{D_k  }{f_{k}^C},
	\end{equation}
	where $f_{k}^C$ is the number of CPU flops allocated by the cloud to execute ${\sf T}_k$ in every second.
	%, and  $\nu_{C}$ is the number of CPU cycles required for each floating point
	%operation at the cloud.
\end{enumerate}

To differentiate the above two cases, we introduce a binary variable $\beta_{k} \in \{0, ~ 1\}$ to indicate where the computation is performed. In particular,
\[
\beta_{k} = \begin{cases}
0, & \mbox{if the 	fog performs computation,} \\
1, & \mbox{if the cloud performs computation.}
\end{cases}\]

%\subsection{Problem Formulation}
Based on the above offloading model,  our goal is to optimize the communication and computation resources,
%, including the UE-Fog association $\alpha_{k,\ell}$, the task distribution $\beta_{k}$, the  beamforming $\bm v_k$, the fronthaul capacity allocation $C_{k,\ell}$ and the CPU flops  $f_{k,\ell}^F$ and $f_{k}^C$, 
so that the maximum latency among UEs is  minimized:\footnote{We consider a situation where the computing outputs contain very few bits, and thus can be delivered to users with negligible time.}
\begin{subequations}\label{eq:main}
	\begin{align}
	& \min_{\substack{  \{\bm v_k, f_{k}^C, \beta_k\}_k, \\ \{f_{k,\ell}^F, C_{k,\ell}, \alpha_{k,\ell}\}_{k,\ell} }}    \max_{k \in {\cal K}}~ \sum_{\ell = 1}^L  \alpha_{k,\ell} \left( \tau_{k,\ell}^T + (1-\beta_{k}) \tau_{k,\ell}^F +   \beta_{k} \tau_{k,\ell}^C  \right)  \notag\\
	& \hspace{1cm} {\rm s.t.}  \quad    f_{k,\ell}^F \leq \alpha_{k,\ell} (1-\beta_{k}) F_{\ell,\max}, ~\forall ~k, \ell, \label{eq:main_b}\\
	& \hspace{1.5cm} ~ \sum_{k=1}^K   f_{k,\ell}^F \leq  F_{\ell,\max}, \quad f_{k,\ell}^F  \geq 0, ~\forall~k, \ell,  \label{eq:main_c}\\
	& \hspace{1.5cm} ~  f_{k}^C \leq     \beta_{k} F_{C,\max} ,  ~ \forall~  k, \label{eq:main_d}\\
	& \hspace{1.5cm} ~ \sum_{k=1}^K  f_{k}^C \leq  F_{C,\max}, \quad f_{k}^C\geq 0, ~\forall~k, \label{eq:main_e}\\
	& \hspace{1.5cm} ~  C_{k,\ell} \leq \alpha_{k,\ell}  \beta_{k} C_{\ell,\max},~ \forall k,~ \ell, \label{eq:main_f}\\
	& \hspace{1.5cm} ~ \sum_{k=1}^K C_{k,\ell} \leq  C_{\ell,\max},~ C_{k,\ell}\geq 0, ~\forall~k,~\ell, \label{eq:main_g}\\
	& \hspace{1.5cm} ~\| \bm v_k\|^2 \leq P_k, ~\forall ~k, \label{eq:main_a}\\
	& \hspace{1.5cm} ~\alpha_{k,\ell} \in \{0, ~ 1\}, \quad \sum_{\ell=1}^L \alpha_{k,\ell} = 1, ~ \forall ~k,  \label{eq:main_h}\\
	& \hspace{1.5cm} ~ \beta_{k} \in \{0, ~ 1\}, ~\forall~k, \label{eq:main_i}
	\end{align}
\end{subequations}
where $F_{\ell, \max}$ and $F_{C, \max}$ are the maximum number of flops that the fog $\ell$ and the cloud can execute in every second, respectively. The constraints \eqref{eq:main_b}-\eqref{eq:main_c} correspond to the computation resource allocation at fog $\ell$. In particular, \eqref{eq:main_b} implies that fog $\ell$ will allocate computing resource for user $k$ only if $\alpha_{k,\ell}=1$ and $\beta_{k}=0$, i.e., user $k$ is associated with fog $\ell$ and meanwhile the task ${\sf T}_k$ is processed at fog $\ell$. Similarly, the constraints \eqref{eq:main_d}-\eqref{eq:main_e} correspond to the computation resource allocation at the cloud. The constraints \eqref{eq:main_f}-\eqref{eq:main_g} are introduced to account for the finite capacity of fronthaul, and \eqref{eq:main_a} limits the peak transmit power at the UEs.

Problem~\eqref{eq:main} is a MINP, which is generally NP-hard. In the next  section, we will develop a tractable approach to~\eqref{eq:main} with a careful treatment of the discrete variables.  

%we show that  problem~\eqref{eq:main} can be reformulated as a discrete-variable-free form, and  thus  continuous optimization algorithms can be employed to handle it.

\section{An Inexact MM Approach to Problem~(5)} \label{sec:approach}
Let us first show that  problem~\eqref{eq:main} can be reformulated as a discrete-variable-free form, and  thus  continuous optimization approach can be leveraged to handle it. Specifically, we have the following result.
\begin{Theorem}\label{theorem:1}
	The MINP problem~\eqref{eq:main} is equivalent to the following continuous optimization problem:
	\begin{subequations}\label{eq:main_reform}
		\begin{align}
		\hspace{-6pt}	\min_{ \substack{  \{\bm v_k, f_{k}^C\}, \\ \{f_{k,\ell}^F, C_{k,\ell}, \theta_{k,\ell}^F, \theta_{k,\ell}^C\} } } &  \max_{k\in {\cal K}}  \sum_{\ell=1}^L  (   \theta_{k,\ell}^F  ( \tau_{k,\ell}^T+\tau_{k,\ell}^F) +    \theta_{k,\ell}^C ( \tau_{k,\ell}^T+\tau_{k,\ell}^C ) ) \label{eq:main_reform_a}\\
		{\rm s.t.} & ~ ~\theta_{k,\ell}^F \geq 0, \quad \theta_{k,\ell}^C \geq 0,   ~\forall ~k, \ell, \label{eq:main_reform_c}\\
		& ~ ~ \sum_{\ell=1}^L  \theta_{k,\ell}^F + \theta_{k,\ell}^C =  1, ~\forall ~k, \label{eq:main_reform_d}\\
		& ~ ~\eqref{eq:main_c}, \eqref{eq:main_e}, \eqref{eq:main_g}~{\rm and}~\eqref{eq:main_a}~{\rm satisfied}.
		\end{align}
	\end{subequations}
\end{Theorem}
\noindent{\it Proof.}~See Appendix~A. \hfill $\blacksquare$

Building upon the above equivalence, we consider solving problem~\eqref{eq:main_reform}. Let us denote
$
\bm \theta_k  = [\theta_{k,1}^F, \ldots, \theta_{k,L}^F, \theta_{k,1}^C, \ldots, \theta_{k,L}^C]^T\in \mathbb{R}^{2L}$ and $
\bm \tau_k    = [\tau_{k,1}^T+\tau_{k,1}^F, \ldots, \tau_{k,L}^T+\tau_{k,L}^F, \tau_{k,1}^T+\tau_{k,1}^C, \ldots, \tau_{k,L}^T+\tau_{k,L}^C]^T \in \mathbb{R}^{2L}.
$
Problem~\eqref{eq:main_reform} is rewritten as
\begin{subequations}\label{eq:main_reform2}
	\begin{align}
	&  \min_{ \substack{  \{\bm v_k, f_{k}^C, \bm \tau_{k}, \bm \theta_{k} \}_k, \\ \{ R_{k,\ell}, f_{k,\ell}^F, C_{k,\ell}\}_{k,\ell} } } ~  \max_{k\in {\cal K}} ~ \bm \theta_k^T \bm \tau_k \label{eq:main_reform2_a0}\\
	& \hspace{0.5cm}  {\rm s.t.}   ~~~~~    R_{k,\ell}  \leq  \phi_{k,\ell}(\bm V),   ~\forall ~k, \ell, \label{eq:main_reform2_a} \\
	& \hspace{0.1cm} ~  \tau_{k,\ell}^T  \geq \frac{B_k}{R_{k,\ell}}, ~  \tau_{k,\ell}^F \geq \frac{D_k  }{f_{k,\ell}^F},~  \tau_{k,\ell}^C \geq \frac{B_k}{C_{k,\ell}} + \frac{D_k  }{f_{k}^C}, \forall~k, \ell, \label{eq:main_reform2_b} \\
	%& \hspace{6.5cm} ~\forall~k, \ell, \label{eq:main_reform2_b}\\
	& \hspace{0.5cm}  ~ \eqref{eq:main_c},~ \eqref{eq:main_e},~ \eqref{eq:main_g},~ \eqref{eq:main_a},~ \eqref{eq:main_reform_c},~\eqref{eq:main_reform_d}~{\rm satisfied}. \label{eq:main_reform2_c}
	\end{align}
\end{subequations}
where $\bm V \triangleq \{\bm v_{k}\}_k$ and $\phi_{k,\ell}(\bm V) \triangleq W \log \Big( 1 +  \bm v_k^H \bm H_{k,\ell} \big(\sigma_\ell^2 \bm I + \sum_{j\neq k}  \bm H_{j,\ell}^H \bm v_j \bm v_j^H \bm H_{j,\ell} \big)^{-1} \bm H_{k,\ell}^H \bm v_k \Big)$. Notice that in \eqref{eq:main_reform2_a} and \eqref{eq:main_reform2_b} we have changed the equalities in \eqref{eq:R_df}-\eqref{eq:tau_C} as inequalities. This does not incur any loss of optimality because the inequalities in \eqref{eq:main_reform2_a} and \eqref{eq:main_reform2_b} must be active at the optimal solution; for otherwise, we can further decrease $\tau_{k,\ell}^X, X\in \{T,F,C\}$ and increase $R_{k,\ell}$ to get a lower objective value.

The constraints \eqref{eq:main_reform2_b} and \eqref{eq:main_reform2_c} are convex, but the objective \eqref{eq:main_reform2_a0} and the constraint \eqref{eq:main_reform2_a} are still nonconvex. For~\eqref{eq:main_reform2_a0}, we handle it by MM. Let  ${\cal X} \triangleq \{  \bm v_k, f_{k}^C, \bm \tau_{k}, \bm \theta_{k}   R_{k,\ell}, f_{k,\ell}^F, C_{k,\ell}\}$ be a collection of optimization variables, and ${\cal F}$ be the feasible set of problem \eqref{eq:main_reform2}. The idea of MM is to  find a surrogate function $g({\cal X}|\bar{\cal X})$, parameterized by some given point $\bar{\cal X} \in {\cal F}$, for the nonconvex objective \eqref{eq:main_reform2_a0} such that the following holds:
\begin{equation}\label{eq:mm_condition}
\begin{aligned}
g({\cal X}|\bar{\cal X}) & \geq \max_{k\in {\cal K}} ~ \bm \theta_k^T \bm \tau_k,\forall~{\cal X}\in {\cal F} , \\
g(\bar{\cal X}|\bar{\cal X}) & = \max_{k\in {\cal K}} ~ \bar{\bm \theta_k}^T \bar{\bm \tau}_k. 
\end{aligned}
\end{equation}
To this end, we make use of the following fact.
\begin{Fact}\label{fact:1} 
	The function
	$$g({\cal X}|\bar{\cal X}) = \max_{k\in \cal K} \left\{   \frac{\| \bm \theta_k + \bm \tau_k\|^2 }{2} - (\bar{\bm \theta}_k)^T \bm \theta_k - (\bar{\bm \tau}_k)^T \bm \tau_k  - \bar{c}_k \right\}$$
	with $\bar{c}_k =  \frac{\| \bar{\bm \theta}_k\|^2 + \| \bar{\bm \tau}_{k} \|^2}{2}$ is a  surrogate function of $\max_{k\in {\cal K}} ~ \bm \theta_k^T \bm \tau_k$. 
\end{Fact}
Fact~\ref{fact:1} can be easily shown by noting  
\begin{equation*}
\begin{aligned}
\bm \theta_k^T \bm \tau_k & = \frac{\| \bm \theta_k + \bm \tau_k\|^2}{2} -  \frac{\| \bm \theta_k\|^2 + \| \bm \tau_k \|^2}{2}\\
&\leq \frac{1}{2} \| \bm \theta_k + \bm \tau_k\|^2 - (\bar{\bm \theta}_k)^T \bm \theta_k - (\bar{\bm \tau}_k)^T \bm \tau_k  - \bar{c}_k 
\end{aligned}
\end{equation*}
for all feasible $(\bar{\bm \theta}_k, \bar{\bm \tau}_k)$, where the inequality is due to the first-order condition for the concave function $-\frac{\| \bm \theta_k\|^2 + \| \bm \tau_k \|^2}{2}$. 
By invoking Fact~\ref{fact:1}, the MM for problem~\eqref{eq:main_reform2} entails repeatedly performing the following updates:
\begin{equation}\label{eq:dc}
\begin{aligned}
{\cal X}^{(t+1)}  =  \arg\min_{\cal X} &   ~ g({\cal X}|{\cal X}^{(t)}) \\
{\rm s.t.} & ~ \eqref{eq:main_reform2_a}-\eqref{eq:main_reform2_c}~{\rm satisfied},
\end{aligned}
\end{equation}
for $t=0,1,2,\ldots$ until some stopping criteria is satisfied.

According to the classical convergence result for MM~\cite{MM}, it is well known that every limit point of the iterates generated by \eqref{eq:dc} is a stationary point of problem~\eqref{eq:main_reform2}. However, this  convergence result holds under the premise that each MM subproblem  is {\it optimally} solved. As for the considered problem~\eqref{eq:dc}, it may be hard to do so due to the nonconvex constraint \eqref{eq:main_reform2_a}. To circumvent this difficulty,  we apply the WMMSE method~\cite{Shi} to find an approximate solution for \eqref{eq:dc}. Specifically,  define by $\bm u_{k,\ell}\in \mathbb{C}^M$ the receive beamformer employed by fog $\ell$ to receive user $k$'s signal. Then, the rate function  $\phi_{k,\ell}(\bm V)$ can be alternatively expressed as~\cite{Shi}:
\begin{equation} \label{eq:rate_wmmse}
\phi_{k,\ell}(\bm V) = \max_{ \bm u_{k,\ell}, w_{k,\ell} \geq 0}    f_{k,\ell}(\bm u_{k,\ell}, w_{k,\ell}, \bm V )
\end{equation}
where 
%$ f_{k,\ell}(\cdot)$ is defined as
\begin{equation}\label{eq:f_def}
f_{k,\ell}(\bm u_{k,\ell}, w_{k,\ell}, \bm V )  \triangleq  W \left( - w_{k,\ell} e_{k,\ell}(\bm u_{k,\ell}, \bm V) + \log w_{k,\ell} + 1 \right)
\end{equation}
and 
\begin{equation*}
\begin{aligned}
& e_{k,\ell}(\bm u_{k,\ell}, \bm V) \\
%& = (1- \bm u_{k,\ell}^H \bm H_{k,\ell}^H \bm v_{k})(1- \bm u_{k,\ell}^H \bm H_{k,\ell}^H \bm v_{k})^H + \bm u_{k,\ell}^H \Big( \sigma_\ell^2 \bm I +  \sum_{j\neq k}  \bm H_{j,\ell}^H \bm v_j \bm v_j^H \bm H_{j,\ell} \Big) \bm u_{k,\ell} \\
\triangleq & \| 1- \bm u_{k,\ell}^H \bm H_{k,\ell}^H \bm v_{k}\|^2 + \textstyle \sum_{j \neq k} \| \bm v_j^H \bm H_{j,\ell}\bm u_{k,\ell} \|^2 +  \sigma_{\ell}^2 \|\bm u_{k,\ell} \|^2
\end{aligned}
\end{equation*}
is the MSE of estimating user $k$'s signal at fog $\ell$, when  the beamformer $\bm u_{k,\ell}$ is used for reception.
By substituting~\eqref{eq:rate_wmmse} into \eqref{eq:dc}, the MM subproblem is equivalently written as
%\begin{equation}\label{eq:dc_eqv}
%\begin{aligned}
%\hspace{-10pt}\min_{{\cal X}, \bm u_{k,\ell}, w_{k,\ell} }~\max_{k \in {\cal K}} & ~ \frac{\| \bm \theta_k + \bm \tau_k\|^2 }{2} - (\bm \theta_k^{(t)})^T \bm \theta_k - (\bm \tau_k^{(t)})^T \bm \tau_k \\
%{\rm s.t.} &  ~ R_{k,\ell}  \leq  f_{k,\ell}(\bm u_{k,\ell}, w_{k,\ell}, \bm V ),   ~\forall ~k, \ell,  \\
%& ~ \eqref{eq:main_reform2_b}-\eqref{eq:main_reform2_c}~{\rm satisfied}.
%\end{aligned}
%\end{equation}
\begin{equation}\label{eq:dc_eqv}
\begin{aligned}
\min_{{\cal X}, \{\bm u_{k,\ell}, w_{k,\ell}\}_{k,\ell} }&~~ g({\cal X}|{\cal X}^{(t)}) \\
{\rm s.t.} &  ~~ R_{k,\ell}  \leq  f_{k,\ell}(\bm u_{k,\ell}, w_{k,\ell}, \bm V ),   ~\forall ~k, \ell,  \\
&~ ~ \eqref{eq:main_reform2_b}-\eqref{eq:main_reform2_c}~{\rm satisfied}.
\end{aligned}
\end{equation}
Problem~\eqref{eq:dc_eqv} can be efficiently handled by block-coordinate descent (BCD) method. In particular, given $\bm V$ the optimal $\bm u_{k,\ell}$ and $ w_{k,\ell}$ for \eqref{eq:dc_eqv}  is given by~\cite{Shi}
\begin{subequations}\label{eq:wmmse_update}
	\begin{align}
	\bm u_{k,\ell} & =    (\sigma_\ell^2 \bm I + \textstyle \sum_{j=1}^K \bm H_{j,\ell}^H  {\bm v}_j  {\bm v}_j^H \bm H_{j,\ell})^{-1} \bm H_{k,\ell}^H \bm v_k, \label{eq:wmmse_update_U}\\
	w_{k,\ell} & = e_{k,\ell}^{-1}(\bm u_{k,\ell}, \bm V). \label{eq:wmmse_update_W}
	\end{align}
\end{subequations}
On the other hand, given $(\bm u_{k,\ell}, w_{k,\ell})$, problem~\eqref{eq:dc_eqv} is convex with respect to the remaining variables, and thus can be optimally solved, say by off-the-shelf software~\texttt{CVX}~\cite{cvx}. Theoretically speaking, the above BCD procedure needs to be performed sufficiently large number of  rounds in order to obtain a good approximate solution  for problem~\eqref{eq:dc}. However, this could incur high computational complexity for each MM update. To trade off the solution quality and the computational complexity, we propose a computationally-cheap inexact MM algorithm for problem~\eqref{eq:main_reform2}; see Algorithm~\ref{algorithm:1}, where for the $t$th MM iteration, we perform only a small number $J^{(t)}$ rounds of BCD update to compute an approximate solution for problem~\eqref{eq:dc}. The parameter $J^{(t)}$ controls the solution quality for each MM iteration. While Algorithm~\ref{algorithm:1} performs MM inexactly, the following theorem reveals that the same convergence result as the exact MM (i.e., using the optimal solution of \eqref{eq:dc} to update  ${\cal X}^{(t+1)}$) holds.
\begin{Theorem}\label{theorem:stationary}
	The iterates $\{ {\cal X}^{(t)} \}_{t=0,1,\ldots}$ generated by Algorithm~\ref{algorithm:1} yield a sequence of non-increasing objective values for problem~\eqref{eq:main_reform2}. Moreover, every limit point of  $\{ {\cal X}^{(t)} \}_{t=0,1,\ldots}$ is a KKT point of problem~\eqref{eq:main_reform2}.
\end{Theorem}
\noindent{\it Proof.}~See Appendix~B. \hfill $\blacksquare$

The idea of proving Theorem~\ref{theorem:stationary} is that the inexact MM (even for the case of $J^{(t)} = 1, \forall~ t$) is sufficient to provide certain improvement for the objective~\eqref{eq:main_reform2_a0}. By accumulating these improvements, the iterations will finally reside at a KKT point of problem~\eqref{eq:main_reform2}. 
%Algorithm~\ref{algorithm:1} summarizes the whole procedure of the proposed inexact BCD approach to problem~\eqref{eq:main_reform2}.
%The detailed proof is omitted due to the page limit.

\begin{algorithm}
	\caption{An Inexact MM Approach to~\eqref{eq:main_reform2}}\label{algorithm:1}
	\begin{algorithmic}[1]
		\State Initialize with a feasible point ${\cal X}^{(0)}$, a set of small positive integers $\{ J^{(t)}\}_{t=0,1,\ldots}$ and set $t=0$
		\Repeat
		\State Set ${\cal X}^{(t_0)}  = {\cal X}^{(t)}$;
		\For {$j=0,1,\ldots,J^{(t)}-1$}
		\State Update $ (\bm u_{k,\ell}^{(t_j)},w_{k,\ell}^{(t_j)})$ according to  \eqref{eq:wmmse_update_U} and \eqref{eq:wmmse_update_W};% with $\bm V = \bm V^{(t_j)}$;
		%\State Update $w_{k,\ell}^{(t_j)}$ according to  \eqref{eq:wmmse_update_W};% with $(\bm V, \bm u_{k,\ell}) = (\bm V^{(t_j)}, \bm u_{k,\ell}^{(t_j)})$;
		\State Update ${\cal X}^{(t_{j+1})}$ by solving~\eqref{eq:dc_eqv} with fixed $(\bm u_{k,\ell}, w_{k,\ell}) = (\bm u_{k,\ell}^{(t_j)}, w_{k,\ell}^{(t_j)})$;
		\EndFor
		\State Set ${\cal X}^{(t+1)} = {\cal X}^{(t_{j+1})}$;
		\State $t \leftarrow t+1$
		\Until{some stopping criterion is satisfied}
		\State {\bf Output} ${\cal X}^{(t)}$.
	\end{algorithmic}
\end{algorithm}

\section{The Cooperative Offloading Case}\label{sec:soft_mode}
%In the last two sections, we have considered a two-stage offloading strategy and developed an inexact BCD algorithm to optimize the UE-Fog association and resource allocations. For such a two-stage offloading strategy,  UE's task is decoded and forwarded to the cloud via the associated fog, if the task is processed at the cloud. However, this decode-and-forward strategy may have some limitations. For example, the fog needs to decode the task information before forwarding, which could increase the complexity at the fog. More importantly, the decode-and-forward strategy may not be able to fully exploit the cooperative gain among fog nodes, because UE's task information is forwarded to the cloud only by the associated fog. In view of this, in this section, we investigate another forwarding strategy, namely, compression and forwarding, where all the fogs will first compress or quantize their received signals, and then cooperatively forward the compressed signals to the cloud.  

In the last two sections, we have considered a two-stage offloading, where each UE's task is  decoded and forwarded to the cloud via the associated fog node, if the task is processed at the cloud. However, this decode-and-forward strategy  may not be able to fully exploit the cooperative gain among the fog nodes.
%, because UE's task information is forwarded to the cloud only by the associated fog. 
%In view of this, 
In this section, we investigate another forwarding strategy, namely, compress-and-forward, where the fog nodes  quantize their received signals  using single-user compression, and then transmit  the compressed bits to the cloud. By doing so, the UEs' signals can be simultaneously delivered to the cloud via all the fog nodes. To put it into context,  recall the received signal model at the fog nodes
\begin{equation}
\bm y_\ell (t) = \sum_{k\in {\cal K} }  \bm H_{k,\ell}^H \bm v_k s_k(t) + \bm n_\ell(t),~\forall~\ell.
\end{equation}
After the reception, the fog $\ell$ quantizes its received signal $\bm y_\ell$. Assuming Gaussian quantization, the quantized signal $\hat{\bm y}_\ell (t) $ is given by
\begin{equation}
\hat{\bm y}_\ell (t) = \bm y_\ell(t) + \bm q_\ell(t),~\ell\in {\cal L},
\end{equation}
where $\bm q_\ell(t)$ is the quantization noise and follows $\bm q_\ell(t) \sim {\cal CN}(\bm 0, \bm Q_\ell)$ with $\bm Q_\ell\succeq \bm 0$~\cite{weiyu16}. Notice that $\bm Q_\ell$ needs to be jointly optimized with the other resource variables to achieve minimum latency.
The quantized signals $\{ \hat{\bm y}_\ell \}_{\ell \in \cal L}$  are then compressed and forwarded to the cloud via the capacity-limited fronthaul links. At the cloud, a two-stage successive decoding strategy is employed---the cloud first recovers the quantized signals $\{ \hat{\bm y}_\ell \}_{\ell \in \cal L}$, and then decodes UEs' messages $\{ s_k\}_{k\in \cal K}$ based on the quantized signals $\{ \hat{\bm y}_\ell \}_{\ell \in \cal L}$. Overall, when 
the compress-and-forward scheme is employed at the fog nodes, the UE-to-cloud channel can be seen as a V-MAC. Following the results in~\cite{weiyu16} and
assuming linear MMSE reception at the cloud, the achievable rate $R_{k,C}$ of UE $k$ for the V-MAC is given by
\begin{equation} \label{eq:rate_soft}
R_{k,C} = W\log(1 + \bm v_k^H \bm H_{k,{\cal L}} \bm J_{k}^{-1} \bm H_{k,{\cal L}}^H \bm v_k)
\end{equation}
where
%\begin{align*}
%\bm H_{k,{\cal L}} & = [\bm H_{k,1}, \ldots, \bm H_{k,L}] \\
%\bm J_{k} & = \sum_{j \neq k} \bm H_{j,{\cal L}}^H \bm v_j \bm v_j^H \bm H_{j,{\cal L}} + \bm \Sigma_{\cal L} + \bm Q_{\cal L} \\
%\bm \Sigma_{\cal L} & = \begin{bmatrix}
%\sigma_1^2 \bm I & & \\
%& \ddots & \\
%& & \sigma_L^2 \bm I
%\end{bmatrix} \\
%\bm Q_{\cal L} & =  \begin{bmatrix}
%\bm Q_1 &  & \\
%& \ddots & \\
%& & \bm Q_L
%\end{bmatrix}.
%\end{align*}
\begin{align*}
\bm H_{k,{\cal L}} & = [\bm H_{k,1}, \ldots, \bm H_{k,L}], \\
\bm J_{k} & = \textstyle \sum_{j \neq k} \bm H_{j,{\cal L}}^H \bm v_j \bm v_j^H \bm H_{j,{\cal L}} + \bm \Sigma_{\cal L} + \bm Q_{\cal L}, \\
\bm \Sigma_{\cal L} & = {\rm Diag}(\sigma_1^2 \bm I, \ldots,\sigma_L^2 \bm I ), \quad 
\bm Q_{\cal L} = {\rm Diag}(\bm Q_1, \ldots, \bm Q_L).
\end{align*}
Since the fog nodes are connected with the cloud via limited-capacity fronthaul, the compression rates at the fog nodes should also satisfy the
fronthaul capacity constraints, so that the cloud can correctly recover the quantized signals $\{ \hat{\bm y}_\ell \}_{\ell \in \cal L}$. Specifically, the fronthaul constraint under single-user compression
is given by
\begin{equation*} \label{eq:soft_mode_fronthaul_constraint}
\log \frac{|\sum_{k=1}^K \bm H_{k,\ell}^H \bm v_k \bm v_k^H \bm H_{k,\ell} + \sigma_\ell^2 \bm I + \bm Q_\ell|}{|\bm Q_\ell|} \leq C_{\ell,\max}, ~ \forall~\ell \in \cal L.
\end{equation*}
If UE $k$'s task is computed at the cloud, the total latency is calculated as
\begin{equation*}
\tau_{k,C} =\frac{B_k}{R_{k,C}}  +  \frac{D_k}{f_{k}^C}.
\end{equation*}

Now, our min-max latency optimization problem under cooperative offloading is formulated as
\begin{subequations}\label{eq:soft_main}
	\begin{align}
	& \min_{\substack{  \{\bm v_k, f_{k}^C, \beta_k\}_k, \\ \{f_{k,\ell}^F, \bm Q_{\ell}, \alpha_{k,\ell}\}_{k,\ell} }}    \max_{k \in {\cal K}}~ \sum_{\ell = 1}^L  \alpha_{k,\ell} \left( (1-\beta_{k}) ( \tau_{k,\ell}^T + \tau_{k,\ell}^F)  +   \beta_{k}  \tau_{k,C}  \right)  \notag\\
	& \hspace{1cm} {\rm s.t.}  \quad    f_{k,\ell}^F \leq \alpha_{k,\ell} (1-\beta_{k}) F_{\ell,\max}, ~\forall ~k, \ell, \label{eq:soft_main_b}\\
	& \hspace{1.5cm} ~ \sum_{k=1}^K   f_{k,\ell}^F \leq  F_{\ell,\max}, ~~ f_{k,\ell}^F  \geq 0, ~\forall~k, \ell,  \label{eq:soft_main_c}\\
	& \hspace{1.5cm} ~  f_{k}^C \leq     \beta_{k} F_{C,\max} ,  ~ \forall~  k, \label{eq:soft_main_d}\\
	& \hspace{1.5cm} ~ \sum_{k=1}^K  f_{k}^C \leq  F_{C,\max}, \quad f_{k}^C\geq 0, ~\forall~k, \label{eq:soft_main_e}\\
	& \hspace{0cm} ~  \log \frac{\left|\displaystyle \sum_{k=1}^K \bm H_{k,\ell}^H \bm v_k \bm v_k^H \bm H_{k,\ell} + \sigma_\ell^2 \bm I + \bm Q_\ell \right|}{|\bm Q_\ell|} \leq C_{\ell, \max}, ~ \forall~\ell, \label{eq:soft_main_f}\\
	& \hspace{1.5cm} ~  \bm Q_\ell \succeq \bm 0, \quad \forall ~\ell \in {\cal L},\label{eq:soft_main_g} \\
	& \hspace{1.5cm} ~\| \bm v_k\|^2 \leq P_k, ~\forall ~k, \label{eq:soft_main_a}\\
	& \hspace{1.5cm} ~\alpha_{k,\ell} \in \{0, ~ 1\}, \quad \textstyle\sum_{\ell=1}^L \alpha_{k,\ell} = 1, ~ \forall ~k,  \label{eq:soft_main_h}\\
	& \hspace{1.5cm} ~ \beta_{k} \in \{0, ~ 1\}, ~\forall~k. \label{eq:soft_main_i}
	\end{align}
\end{subequations}
%Compared with problem~\eqref{eq:main}, problem~\eqref{eq:soft_main} involves the additional compression variables $\{ \bm Q_\ell \}_{\ell \in \cal L}$  and  models the fronthaul capacity constraint from an information theoretic viewpoint. 

Similar to problem~\eqref{eq:main}, problem~\eqref{eq:soft_main} is an MINP. Following the proof of Theorem~\ref{theorem:1}, one can show that  problem~\eqref{eq:soft_main} can be reformulated as the following discrete-variable-free form:
\begin{subequations}\label{eq:soft_main_eqv}
	\begin{align}
	& \min_{\substack{  \{\bm v_k, f_{k}^C\}_k, \\ \{f_{k,\ell}^F, \bm Q_{\ell}, \theta_{k,\ell}^F, \theta_{k,C}  \}}}    \max_{k \in {\cal K}}~  \left\{ \theta_{k,C} \tau_{k,C} + \sum_{\ell = 1}^L  \theta_{k,\ell}^F ( \tau_{k,\ell}^T + \tau_{k,\ell}^F) \right\}  \notag\\
	& \hspace{1cm} {\rm s.t.}  \quad    \theta_{k,\ell}^F \geq 0,~\forall ~k, \ell, \quad \theta_{k,C}\geq 0,~\forall~k, \label{eq:soft_main_a_eqv}\\
	& \hspace{1.5cm} ~ \theta_{k,C} + \textstyle \sum_{\ell=1}^L \theta_{k,\ell} =1,  \label{eq:soft_main_b_eqv}\\
	& \hspace{1.5cm} ~ \eqref{eq:soft_main_c}, \eqref{eq:soft_main_e}-\eqref{eq:soft_main_a}~{\rm satisfied}.
	\end{align}
\end{subequations}
Denote $\bm \theta_k = [\theta_{k,C}, ~\theta_{k,1}, \ldots, \theta_{k,L}] \in \mathbb{R}^{L+1}$ and $\bm \tau_k = [\tau_{k,C}, ~ \tau_{k,1}^T + \tau_{k,1}^F, \ldots, \tau_{k,L}^T + \tau_{k,L}^F ] \in \mathbb{R}^{L+1}$. Problem~\eqref{eq:soft_main_eqv} can be reexpressed as
\begin{subequations}\label{eq:soft_main_eqv2}
	\begin{align}
	& \min_{\substack{  \{\bm v_k, f_{k}^C, \bm \theta_k, \bm \tau_k\}_k, \\ \{f_{k,\ell}^F, \bm Q_{\ell} \}}}    \max_{k \in {\cal K}}~  \bm \theta_k^T \bm \tau_k \notag\\
	& \hspace{0cm} {\rm s.t.}  ~  R_{k,\ell} \leq  W \log  \Big( 1 +  \bm v_k^H \bm H_{k,\ell} \times  \notag\\
	&\hspace{1.3cm}  (\sigma_\ell^2 \bm I + \sum_{j\neq k}  \bm H_{j,\ell}^H \bm v_j \bm v_j^H \bm H_{j,\ell}  )^{-1} \bm H_{k,\ell}^H \bm v_k  \Big)  \label{eq:soft_main_a_eqv2}\\
	& \hspace{.5cm}~  R_{k,C} \leq  W\log(1 + \bm v_k^H \bm H_{k,{\cal L}} \bm J_{k}^{-1} \bm H_{k,{\cal L}}^H \bm v_k) \label{eq:soft_main_b_eqv2}\\
	& \hspace{0.cm}   \tau_{k,\ell}^T  \geq \frac{B_k}{R_{k,\ell}},  \,  \tau_{k,\ell}^F \geq \frac{D_k  }{f_{k,\ell}^F}, \, \tau_{k,C} \geq \frac{B_k}{R_{k,C}} + \frac{D_k  }{f_{k}^C}, \forall~k,\ell,  \label{eq:soft_main_c_eqv2}\\
	& \hspace{0.cm}  \log \frac{ \left| \displaystyle \sum_{k=1}^K \bm H_{k,\ell}^H \bm v_k \bm v_k^H \bm H_{k,\ell} + \sigma_\ell^2 \bm I + \bm Q_\ell \right|}{|\bm Q_\ell|} \leq C_{\ell,\max},   \forall~\ell \label{eq:soft_main_d_eqv2} \\
	%	&  \hspace{6.1cm}  ~ \forall~\ell \in \cal L, \label{eq:soft_main_d_eqv2}\\
	& \hspace{.5cm} ~ \eqref{eq:soft_main_a_eqv}-\eqref{eq:soft_main_b_eqv},  \eqref{eq:soft_main_c},  \eqref{eq:soft_main_e},  \eqref{eq:soft_main_g},  \eqref{eq:soft_main_a}~{\rm satisfied}.
	\end{align}
\end{subequations}

The constraints~\eqref{eq:soft_main_a_eqv2}-\eqref{eq:soft_main_b_eqv2} can be handled similarly as before by using WMMSE reformulation. Specifically, the constraints~\eqref{eq:soft_main_a_eqv2} and \eqref{eq:soft_main_b_eqv2} can be expressed as
\begin{equation}\label{eq:R_k_reform}
R_{k,\ell}  \leq  f_{k,\ell}(\bm u_{k,\ell}, w_{k,\ell}, \bm V ) 
\end{equation} 
and
\begin{equation} \label{eq:R_C_reform}
R_{k,C} \leq  f_{k,C}(\bm u_{k,C}, w_{k,C}, \bm V, \bm Q_{\cal L} ),
\end{equation} 
respectively, where $f_{k,\ell}$ is defined in~\eqref{eq:f_def},
%\begin{align*}
%&f_{k,C}(\bm u_{k,C}, w_{k,C}, \bm V, \bm Q_{\cal L} )   \\
%\triangleq  & W \left( - w_{k,C} e_{k,C}(\bm u_{k,C}, \bm V, \bm Q_{\cal L}) + \log(w_{k,C}) + 1 \right), \\
%& e_{k,C}(\bm u_{k,C}, \bm V, \bm Q_{\cal L})  \\
%\triangleq&  \| 1- \bm u_{k,C}^H \bm H_{k,{\cal L}} \bm v_{k}\|^2 + \textstyle \sum_{j \neq k} \|\bm u_{k,C}^H \bm H_{j,{\cal L}}  \bm v_j \|^2 \\
%& +    \bm u_{k,C}^H (\bm \Sigma_{\cal L} + \bm Q_{\cal L}) \bm u_{k,C}.
%\end{align*} 
$f_{k,C}(\bm u_{k,C}, w_{k,C}, \bm V, \bm Q_{\cal L} )  
\triangleq W \left( - w_{k,C} e_{k,C}(\bm u_{k,C}, \bm V, \bm Q_{\cal L}) + \log(w_{k,C}) + 1 \right)$, and $e_{k,C}(\bm u_{k,C}, \bm V, \bm Q_{\cal L})  
\triangleq  \| 1- \bm u_{k,C}^H \bm H_{k,{\cal L}} \bm v_{k}\|^2 + \textstyle \sum_{j \neq k} \|\bm u_{k,C}^H \bm H_{j,{\cal L}}  \bm v_j \|^2 +    \bm u_{k,C}^H (\bm \Sigma_{\cal L} + \bm Q_{\cal L}) \bm u_{k,C}$.

As for the fronthaul-capacity constraint~\eqref{eq:soft_main_d_eqv2}, the following lemma is leveraged to recast it into a more tractable form:
\begin{Lemma}[\cite{qli_jsac}~]\label{lemma:1}
	Let ${\bf E} \in \mathbb{C}^{N \times N}$
	be any matrix such that ${\bf E} \succ {\bf 0}$. Consider the
	function $f({\bf S}) = - {\rm Tr} ({\bf S} {\bf E}) +  \ln | {\bf S} | + N
	$. Then,
	\begin{equation} \label{eq:conjugate_lemma}
	\ln | {\bf E}^{-1} |  = \max_{{\bf S} \in \mathbb{C}^{N \times N}, {\bf S} \succeq {\bf 0}} f({\bf S}),
	\end{equation}
	and the optimal ${\bf S}^\star = {\bf E}^{-1}$.
\end{Lemma}
Applying Lemma~\ref{lemma:1} to the constraint~\eqref{eq:soft_main_d_eqv2} yields
\begin{equation}\label{eq:log_det_reform}
\max_{\bm S_\ell \succeq \bm 0} \{ -{\rm Tr} (\bm S_\ell \bm E_\ell) + \log |\bm S_\ell| + M_\ell\} + \log|\bm Q_\ell| + C_{\ell,\max} \geq 0, ~\forall ~\ell
\end{equation}
where $\bm E_\ell = \sum_{k=1}^K \bm H_{k,\ell}^H \bm v_k \bm v_k^H \bm H_{k,\ell} + \sigma_\ell^2 \bm I + \bm Q_\ell$.
By substituting~\eqref{eq:R_k_reform}, \eqref{eq:R_C_reform} and \eqref{eq:log_det_reform} into \eqref{eq:soft_main_a_eqv2}, \eqref{eq:soft_main_b_eqv2} and \eqref{eq:soft_main_d_eqv2}, respectively,  we can equivalently express problem~\eqref{eq:soft_main_eqv2} as
\begin{equation}\label{eq:soft_main_eqv3}
\begin{aligned}
& \min_{\substack{  \{\bm v_k, \bm u_k, \bm w_k, f_{k}^C, \bm \theta_k, \bm \tau_k\}_{k,\ell}, \\ \{f_{k,\ell}^F, \bm Q_{\ell}, \bm S_\ell, R_{k,\ell} \}_{k,\ell}}}    \max_{k \in {\cal K}}~  \bm \theta_k^T \bm \tau_k \\
& \hspace{-.1cm} {\rm s.t.}  ~~  R_{k,\ell}  \leq  f_{k,\ell}(\bm u_{k,\ell}, w_{k,\ell}, \bm V ) ,  \\
& \hspace{.5cm}~  R_{k,C} \leq  f_{k,C}(\bm u_{k,C}, w_{k,C}, \bm V, \bm Q_{\cal L} ),  \\
& \hspace{.5cm} ~  \tau_{k,\ell}^T  \geq \frac{B_k}{R_{k,\ell}}, ~~  \tau_{k,\ell}^F \geq \frac{D_k  }{f_{k,\ell}^F},~~  \tau_{k,C} \geq \frac{B_k}{R_{k,C}} + \frac{D_k  }{f_{k}^C},   \\
& \hspace{-0.2cm}    -{\rm Tr} (\bm S_\ell \bm E_\ell) + \log |\bm S_\ell| + M_\ell  + \log|\bm Q_\ell| + C_{\ell,\max} \geq 0, \\
& \hspace{.5cm} ~ \bm S_\ell \succeq \bm 0,~ \forall~\ell, \\
& \hspace{.5cm} ~ \eqref{eq:soft_main_a_eqv}-\eqref{eq:soft_main_b_eqv},  \eqref{eq:soft_main_c},  \eqref{eq:soft_main_e},  \eqref{eq:soft_main_g},  \eqref{eq:soft_main_a}~{\rm satisfied}.
\end{aligned}
\end{equation}
Let us denote $\tilde{\cal X} \triangleq \{ \bm \theta_k, \bm \tau_k, \bm V, f_k^C, f_{k,\ell}^F, \bm Q_\ell \}_{k,\ell}$. Notice that by fixing $ \{ \bm u_{k,\ell}, w_{k,\ell}, \bm S_\ell \}_{k,\ell}$ in \eqref{eq:soft_main_eqv3}, the feasible set of problem~\eqref{eq:soft_main_eqv3} is convex with respect to $\tilde{\cal X}$. Meanwhile, given $\tilde{\cal X}$ the optimal  $ \{ \bm u_{k,\ell}, w_{k,\ell}, \bm S_\ell \}_{k,\ell}$ for problem~\eqref{eq:soft_main_eqv3} can be computed in closed form by \eqref{eq:wmmse_update} and Lemma~\ref{lemma:1}. Therefore, problem~\eqref{eq:soft_main_eqv3} can be handled similarly as before by using the MM and the BCD method; the detailed procedure is summarized in Algorithm~\ref{algorithm:2}. Moreover, following a similar proof of Theorem~\ref{theorem:stationary}, it can be shown that every limit point generated by Algorithm~\ref{algorithm:2} is a KKT point of problem~\eqref{eq:soft_main_eqv2}. We omit the detailed proof for brevity.

\begin{algorithm}
	\caption{An Inexact MM Approach to~\eqref{eq:soft_main_eqv3}}\label{algorithm:2}
	\begin{algorithmic}[1]
		\State Initialize with a feasible point $\tilde{\cal X}^{(0)}$, a set of small positive integers $\{ J^{(t)}\}_{t=0,1,\ldots}$ and set $t=0$
		\Repeat
		\State Set $ \tilde{\cal X}^{(t_0)}  = \tilde{\cal X}^{(t)}$;
		\For {$j=0,1,\ldots,J^{(t)}-1$}
		\State Update $(\bm u_{k,\ell}^{(t_{j})}, w_{k,\ell}^{(t_{j})} )$ according to \eqref{eq:wmmse_update_U}  and  \eqref{eq:wmmse_update_W};
		\State Update $\bm S_\ell^{(t_{j})}=( \bm E_\ell^{(t_{j})})^{-1},~\forall~\ell$;
		\State Update ${\cal X}^{(t_{j+1})}$ by solving problem~\eqref{eq:soft_main_eqv3} with  $(\bm S_\ell^{(t_j)}, \bm u_{k,\ell}^{(t_j)}, w_{k,\ell}^{(t_j)})_{k,\ell}$ fixed and the objective replaced by its majorant $g(\tilde{\cal X}|\tilde{\cal X}^{(t)})$;
		\EndFor
		\State Set $\tilde{\cal X}^{(t+1)} = \tilde{\cal X}^{(t_{j+1})}$;
		\State $t \leftarrow t+1$
		\Until{some stopping criterion is satisfied}
		\State {\bf Output} $\tilde{\cal X}^{(t)}$.
	\end{algorithmic}
\end{algorithm}

%\newpage

\section{Simulation Results}\label{sec:sim}
In this section, we test the performance of the proposed offloading schemes by Monte-Carlo simulations. The following simulation settings are used, unless otherwise specified: all the UEs have the same number of transmit antennas $N_j =4,~\forall~j\in{\cal K}$; all the fog nodes have the  same number of receive antennas $ M_{\ell} = 8, ~\forall~\ell \in{\cal L}$; the maximum transmit power at the  $k$th UE is $P_k = 30$\,dBm, $\forall~k\in {\cal K}$, the wireless transmission bandwidth is $W=20$\,MHz and  the noise's variances is normalized to one. For simplicity, we set $J^{(t)}=1, ~\forall~t$  in Algorithm~\ref{algorithm:1}. We consider that there are $L=4$ fog nodes and $K=10$ UEs, which are randomly distributed in the cell with radius $1\times 10^3$\,m. The channels were randomly generated according to the distance model--- the channel coefficients between user $k$ and fog $\ell$ are modeled as zero mean circularly
symmetric complex Gaussian vector with $(2000/d_{k,\ell})^3 \beta_{k,\ell}$ as variance for both real and imaginary dimensions, where $10\log10(\beta_{k,\ell}) \sim {\cal N}(0, 64)$ is a real Gaussian random variable modeling the shadowing effect. In the ensuring two subsections, we will first study the performance of non-cooperative offloading  in Section~\ref{sec:model}-\ref{sec:approach}, and then the cooperative offloading  in Section~\ref{sec:soft_mode}.

\subsection{The Non-cooperative Offloading Case}
\begin{figure}[!h]
	\centerline{\resizebox{.5\textwidth}{!}{\includegraphics{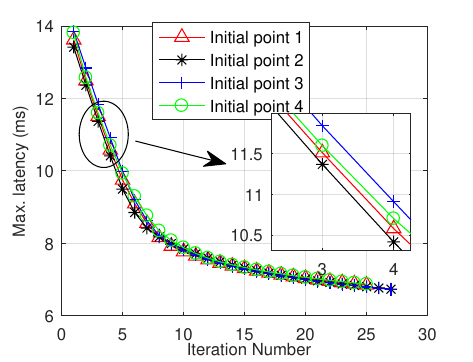}}
	}
	%	\vspace{-5pt}
	\caption{Convergence behavior of Algorithm~\ref{algorithm:1} with different initializations.} \label{fig:example1_a}
\end{figure}

In the first example, we investigate the convergence behavior of Algorithm~\ref{algorithm:1}. We  set $F_{C,\max}=2\times10^3$ (Gflops/sec), $F_{1:4,\max} =[3,~ 4,~ 4,~ 5]\times 10^2$ (Gflops/sec), $C_{1:4,\max} = [30,~ 35,~ 40,~ 50]$ (Mbps),  $D_{1:10} = [2, 2, 2, 6, 6, 6, 6, 8, 8, 8]\times 10^2$ (Mflops) and $B_{1:10}=[20, 20, 20, 40, 40, 40, 40, 60, 60, 60]$ (Kbits). Figure~\ref{fig:example1_a} shows the result, where four random initializations are tested. From the figure, we have the following observations. First,  the maximum latency  decreases monotonically as the iteration number increases. In particular, after 20 iterations, the maximum latency has already decreased from 14\,ms to 7\,ms, and all the tests converge after 25 iterations. This validates the conclusion  in Theorem~\ref{theorem:stationary}. Secondly,  different initializations lead to almost the same convergence  process with the same convergence rate and convergent  value, which demonstrates that Algorithm~\ref{algorithm:1} is not  sensitive to the initialization. 
This  property is favorable when non-convex optimization problem is considered.

\begin{figure}[!h]
	\centerline{\resizebox{.5\textwidth}{!}{\includegraphics{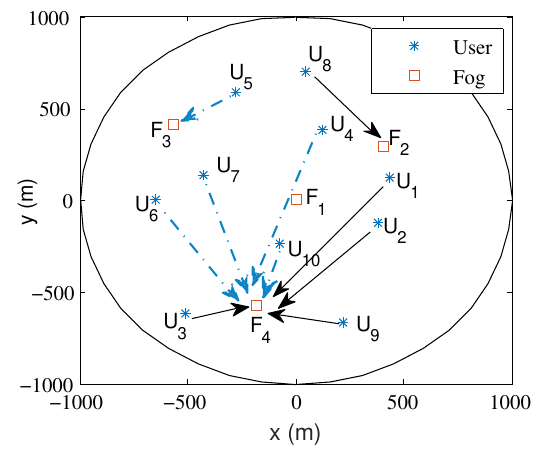}}
	}
	%	\vspace{-5pt}
	\caption{UE-Fog association after convergence.} \label{fig:example1_b}
\end{figure}

Figure~\ref{fig:example1_b} shows the corresponding UE-Fog association and the task distribution after convergence in Figure~\ref{fig:example1_a}. The arrow, which starts  from UE and ends up at fog node, means that the UE offloads its task via the connected fog node. In particular, the solid black line means that the computation is performed at the fog node, and the blue broken line means that  the computation is done at the cloud. From the figure we have the following observations: First, the UE-Fog association is not solely determined by the distance; i.e., UEs may offload their tasks to the fog nodes with larger distance. For example, most of UEs offload  tasks via the fourth fog node, because the fourth fog node has the most powerful communication and computation capability. Therefore, Algorithm~\ref{algorithm:1} can adaptively assign the UE-Fog association according to the available communication and  computation resources. Secondly, when looking into the UEs connected with the  fourth  fog node, we found that the fourth fog node tends to locally perform the computation for those UEs with relatively ``easy'' tasks, i.e., smaller $D_k$ and $B_k$ such as UE1, UE2 and UE3, and forward the ``hard'' tasks to the cloud such as UE4, UE6 and UE7. This is intuitively reasonable, since the latency of completing the hard tasks is dominated by the computation latency rather than the transmission latency.

\begin{figure}[!h]
	\centerline{\resizebox{.5\textwidth}{!}{\includegraphics{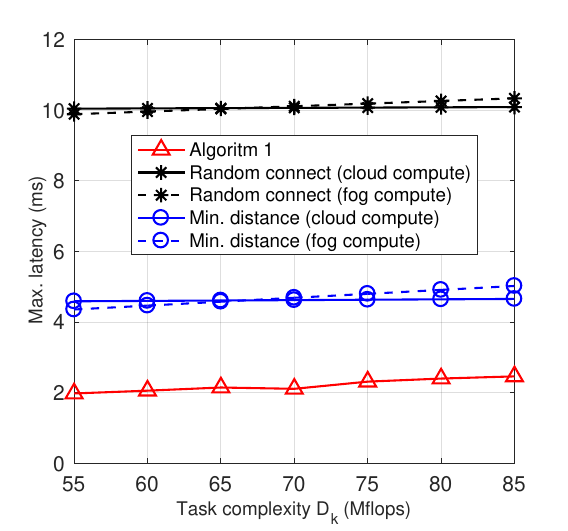}}
	}
	%	\vspace{-5pt}
	\caption{Maximum latency vs. $D_k$ (Mflops).} \label{fig:example4_hd}
\end{figure}

In the second example, we study how the task's complexity $D_k$ affects the latency.  For simplicity, we assume that all the fog nodes have the same computation capacity $F_{\ell,\max}=200$\,(Gflops/second) and the same fronthaul capacity $C_{\ell, \max} = 200$\,(Mbps); all the UEs have the same $B_k=60$\,(Kbits) and $D_k$; the cloud's computation capacity is $F_{C,\max}=2\times10^3$\,(Gflops/second). For comparison, we  have  included two heuristic UE-Fog association strategies, namely, the minimum distance-based association and the random association, under which all the tasks are  offloaded to the connected fog nodes or the cloud, and the fog nodes or the cloud equally allocate  their  resources for the served UEs. 
%In addition, to demonstrate the effectiveness of the joint optimization of UE-Fog association and the task distribution, we have also considered a variant of Algorithm~\ref{algorithm:1} with prefixing $\beta_{k}=1$; i.e., all the tasks are offloaded to the cloud, and only the UE-Fog association and the computation resource allocation  at the cloud are optimized. We name this scheme as ``Alg.~1 (all cloud)'' in the figure. 
The result is shown in Figure~\ref{fig:example4_hd}. From the figure, we see that the proposed Algorithm~\ref{algorithm:1} attains the minimum latency among the compared methods. The minimum distance-based offloading strategy is better than the random one, but there is still a notable performance gap between the former and Algorithm~\ref{algorithm:1}. In addition, we see that for both random connection and minimum distance-based connection schemes, the fog node computation is slightly better than the cloud computation, because  for low-complexity tasks the latency is dominated by the transmission latency and further offloading to the cloud could incur larger latency. Actually, for the proposed Algorithm~\ref{algorithm:1} we see a similar trend. More specifically, we calculate the ratio of tasks that are computed at the fog nodes for Algorithm~\ref{algorithm:1} under the same setting as Figure~\ref{fig:example4_hd}, and the result is shown in Table~\ref{table:task_at_fog_Dk}. As seen,  with the increase of the tasks' complexity,   Algorithm~\ref{algorithm:1}  adaptively assigns more tasks to the cloud.  

%In addition, we see that when $D_{k}$ is small, Algorithm~\ref{algorithm:1}  is better than ```Alg.~1 (all cloud)'', which implies that for light task, it would be preferable to process it at the fog nodes. To verify this, we have also checked the ratio of the tasks completed at the fog nodes; the result is Tabulated in Table~\ref{table:task_at_fog_Dk}. As expected, Algorithm~\ref{algorithm:1} can adaptively adjust the number of tasks processed at the fog nodes according to the complexity of the tasks.  

\begin{table}
	%	\vspace{-.5em}
	\centering
	\caption{Ratio of tasks computed at the fog nodes vs. $D_k$.}
	\label{table:task_at_fog_Dk}
	{  \begin{tabular}{|c|c|c|c|c|c|c|c|}
			\hline
			$D_k$ (Mflops) 	   &  55 & 60 &65 &70 & 75 & 80 & 85 \\\hline
			%		Ratio (\%) &  97\% & 91\% & 79\% & 55\% & 30\% & 20\% & 17\% 
			Ratio (\%) &  97  & 91  & 79 & 55  & 30  & 20  & 17 
			\\\hline
	\end{tabular}}
	%\vspace*{-.1em}
\end{table}

\begin{figure}[!h]
	\centerline{\resizebox{.5\textwidth}{!}{\includegraphics{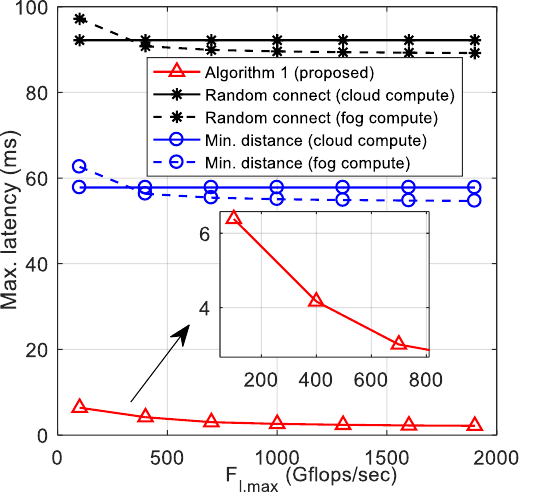}}
	}
	%	\vspace{-5pt}
	\caption{Maximum latency vs. fog nodes' computation capacity $F_{\ell,\max}$.} \label{fig:example3_hd}
\end{figure}

In the third example, we study how the maximum latency changes with the increase of the fog nodes' computation capacity $F_{\ell,\max}$. The simulation is basically the same as the last one, except that we increase $B_k=150$\,(Kbits), and $D_k=200$\,(Mflops). The result is shown in Figure~\ref{fig:example3_hd}.  As expected, when the fog nodes' computation capacity increases, the maximum latency of all the schemes decrease. In addition, for the two compared schemes, they prefer  to perform the computation at the fog nodes when $F_{\ell,\max}$ exceeds  400\,Gflops/sec, since in such a case the latency is dominated by the transmission.  Similar to Figure~\ref{fig:example4_hd},  the performance of Algorithm~\ref{algorithm:1} is still far better than the other two schemes for all the tested $F_{\ell,\max}$.

%Moreover, when $F_{\ell,\max}$ is small, Algorithm~\ref{algorithm:1}  almost coincides with the ``Alg.~1 (all cloud)''; this is reasonable, because in such a case, the bottleneck lies in the computation aspect, and to speed up the computation, the tasks should be processed at the powerful cloud. However, with the improvement of the fog nodes' computation capacity,  more fog nodes should participate in computation to further reduce the latency.

\begin{figure}[!h]
	\centerline{\resizebox{.55\textwidth}{!}{\includegraphics{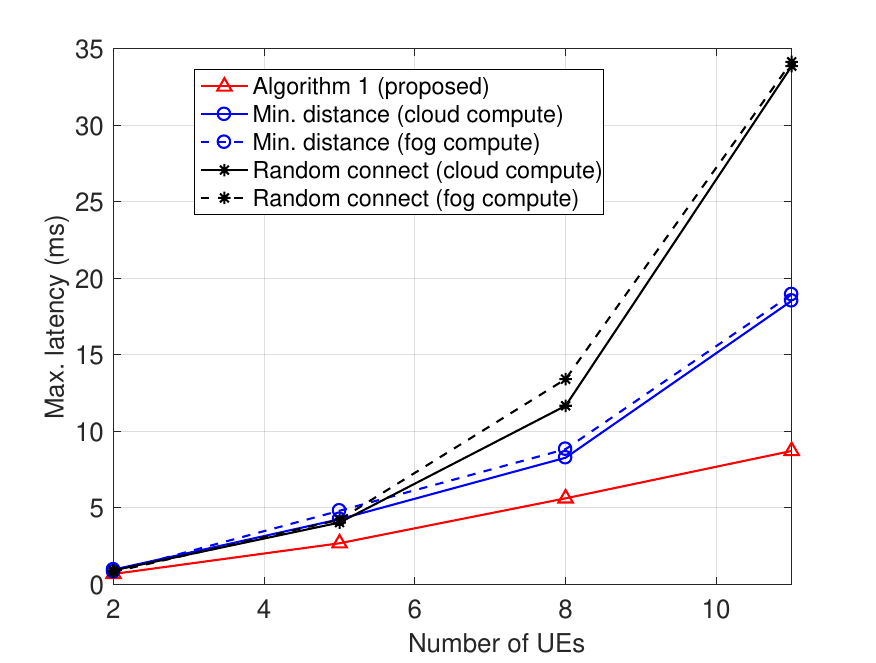}}
	}
	%	\vspace{-5pt}
	\caption{Maximum latency vs. number of UEs.} \label{fig:example2}
\end{figure}

In the fourth example, we investigate the relationship between the number of users and the maximum latency for different offloading strategies. The number of users increases from 2 to 11 according to the setting in Figure~\ref{fig:example1_a}, and the result is shown in Figure~\ref{fig:example2}. We see that with the increase of UEs, the maximum latency of all the schemes increases, but at different speed. Particularly, the random association scheme is more sensitive to the number of UEs, due to the lack of optimization for the UE-Fog association. Also, the proposed  Algorithm~\ref{algorithm:1} yields the best performance among the compared offloading schemes.

\subsection{The Cooperative Offloading Case}
\begin{figure}[!h]
	\centerline{\resizebox{.5\textwidth}{!}{\includegraphics{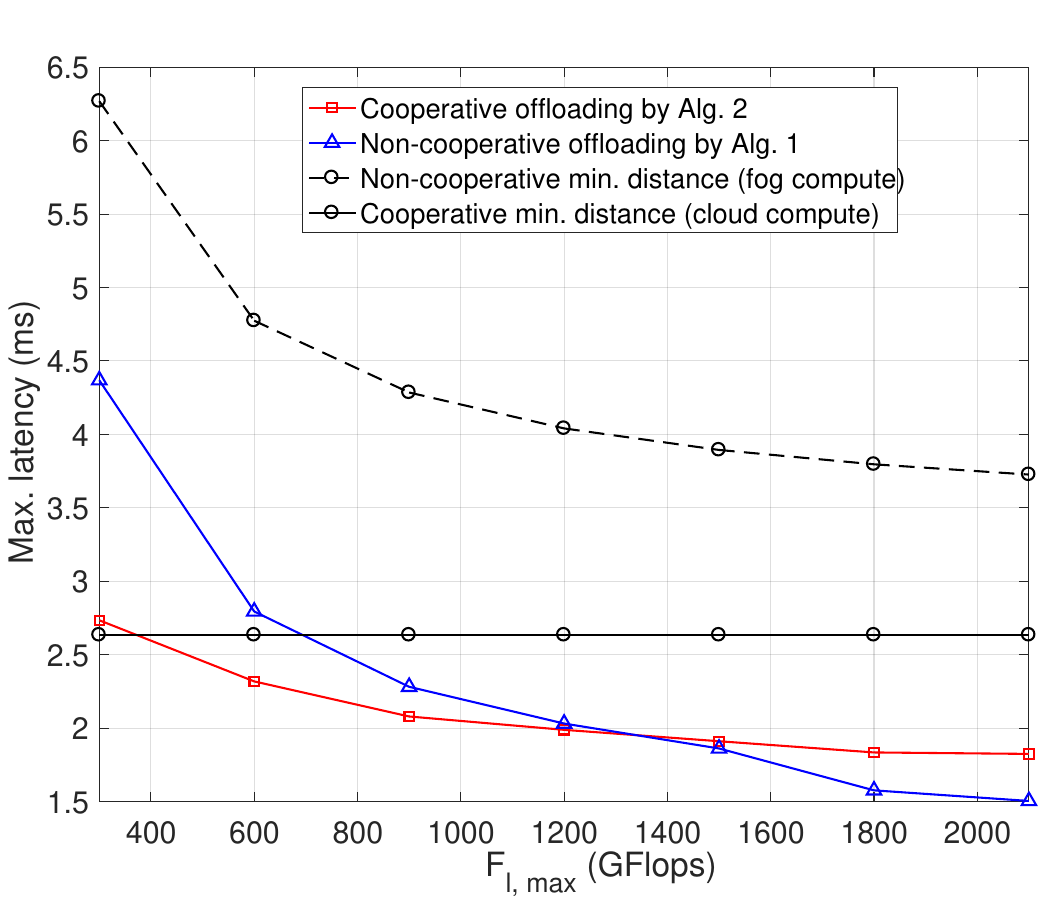}}
	}
	%	\vspace{-5pt}
	\caption{Maximum latency vs. fog nodes' maximum computation capacity $F_{\ell,\max}$ for cooperative and non-cooperative offloading.} \label{fig:example2_soft}
\end{figure}

In this subsection, we study the performance of the cooperative offloading scheme, and make a comparison with the previous non-cooperative offloading. 
%In the first example, 
%let us check the convergence of the proposed Algorithm~\ref{algorithm:2}. The simulation setting is identical to Figure~\ref{fig:example1_a}, and the result is shown in Figure~\ref{fig:example1_soft}. Similar to Figure~\ref{fig:example1_a}, Algorithm~\ref{algorithm:2} yields a monotonically decreasing objective values, and it is also not sensitive to the initialization. 
%Moreover, as compared with Figure~\ref{fig:example1_a}.
%In the second next example, 
In the first example, we compare the performance of the cooperative and non-cooperative offloading schemes, when the fog nodes' computation capacity $F_{\ell,\max}$ increases. For simplicity, we assume that all the fog nodes have the same $F_{\ell,\max}$ and $F_{C,\max} = 1.5\times 10^3$\, (Gflops/sec), $C_{\ell,\max}  = 200$\,(Mbps), $\forall~\ell\in {\cal L}$, $B_k =100$\,(Kbits) and $D_k=200$\,(Mflops),~$\forall~k\in {\cal K}$. The result is shown in Figure~\ref{fig:example2_soft}. From the figure, we see that with the increase of $F_{\ell,\max}$, the maximum latency decreases consistently. In particular, for small-to-medium $F_{\ell,\max}$ the cooperative offloading attains smaller latency than the non-cooperative one. However, when the fog nodes' computation capacity exceeds the cloud's, i.e.,  $F_{\ell,\max} \geq 1.5\times 10^3$\,(Gflops/sec), the non-cooperative offloading becomes better. This can be explained as follows: When fog nodes have sufficient computation resources, it would be more preferable to process the tasks at the fog nodes, rather than compress-and-forwarding the tasks to the cloud, because the latter may further incur latency due to the capacity-limited fronthaul links. To verify this, we tabulate  the ratio of tasks computed at the fog nodes for the two offloading schemes in Table~\ref{table:task_at_fog_F_L}. It can be seen that the non-cooperative offloading has more fog nodes participating in the computation. By contrast, the cooperative scheme tends to offload the task to the cloud, because the  cooperative compress-and-forwarding  can better exploit the transmission diversity  to reduce the transmission latency as compared with the non-cooperative offloading. In addition, we see that the ``Cooperative min. distance (cloud compute)'' scheme is better than the ``non-cooperative min. distance (fog compute)'' scheme, even if the fog's computation capability exceeds the cloud's. This again demonstrates the advantage of cooperative transmission in reducing the transmission latency.

%\begin{figure*}[!t]
%\begin{table}
%	%	\vspace{-.5em}
%%	\centering
%	\caption{Ratio of tasks computed at the fog nodes vs. fog nodes' computation capacity $F_{\ell,\max}$.}
%	\label{table:task_at_fog_F_L}
%	\begin{tabular}{|c|c|c|c|c|c|c|c|}
%		\hline
%		$\substack{F_{\ell,\max}\\ (\times 100\, {\rm Gflops/sec})}$  &  3 & 6 & 9 &12 & 15 & 18 & 21 \\\hline
%		$\substack{{\rm Non-cooperative}\\(\%)}$ &  90 & 100 & 100 & 100 & 100 & 100 & 100 \\\hline
%		$\substack{{\rm Cooperative}\\(\%)}$ &  21 & 37 & 50 & 61 & 73 & 76 & 79  
%		\\\hline
%	\end{tabular}
%	%\vspace*{-.5em}
%\end{table}
%\hrulefill
%\end{figure*}

\begin{table}
	%	\vspace{-.5em}
	\centering
	\caption{Ratio of tasks computed at the fog nodes vs. fog nodes' computation capacity $F_{\ell,\max}$.}
	\label{table:task_at_fog_F_L}
	{ \begin{tabular}{|c|c|c|c|c|c|c|c|}
			\hline
			$\substack{F_{\ell,\max}\\ (\times 100\, {\rm Gflops/sec})}$  &  3 & 6 & 9 &12 & 15 & 18 & 21 \\\hline
			$\substack{{\rm Non-cooperative}\\(\%)}$ &  90 & 100 & 100 & 100 & 100 & 100 & 100 \\\hline
			%		{\rm Non-cooperative (\%)}  &  90 & 100 & 100 & 100 & 100 & 100 & 100 \\\hline
			$\substack{{\rm Cooperative}\\(\%)}$ &  21 & 37 & 50 & 61 & 73 & 76 & 79  
			\\\hline
	\end{tabular}}
	%\vspace*{-.5em}
\end{table}

\begin{figure}[!h]
	\centerline{\resizebox{.5\textwidth}{!}{\includegraphics{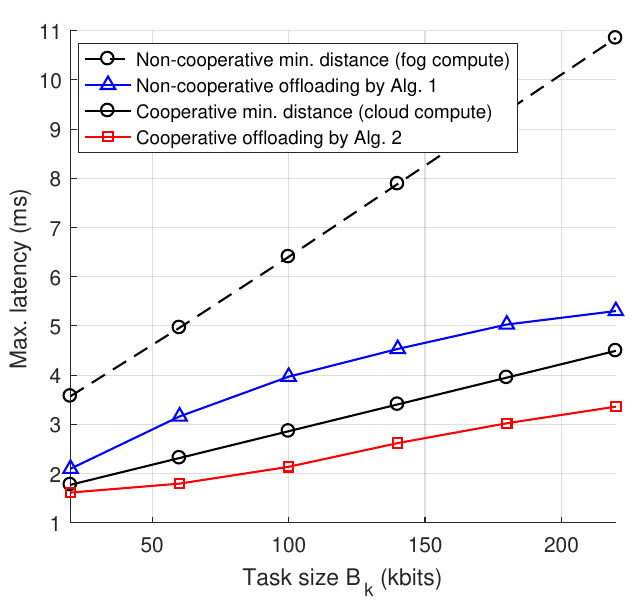}}
	}
	%	\vspace{-5pt}
	\caption{Maximum latency vs. the task size $B_k$.} \label{fig:example4_soft}
\end{figure}

In the second example, we investigate the effect of the task size $B_k$ on the latency. We assume that all the UEs have same $B_k$, and other simulation parameters are $F_{\ell,\max}=500$\, (Gflops/sec), $F_{C,\max}=2\times 10^3$\,(Gflops/sec), $C_{\ell,\max}  = 200$\,(Mbps), $\forall~\ell\in {\cal L}$ and $D_k=300$\,(Mflops),~$\forall~k\in {\cal K}$. The result is shown in Figure~\ref{fig:example4_soft}. As expected, the latency increases with $B_k$, and  the cooperative offloading is consistently better than the non-cooperative one for all the tested $B_k$. Interestingly, under the cooperative mode, even the minimum distance-based UE-Fog association scheme can outperform the non-cooperative offloading; similar observation can be seen in Figure~\ref{fig:example2_soft} for $F_{\ell,\max}\leq 600$\,(Gflops/sec). This demonstrates that the cooperative gain is important for reducing latency.  

%Table~\ref{table:task_at_fog_Bk} lists the ratio of tasks computed  at the fog nodes. 

%{\color{blue}Interestingly, for $B_k \leq 40$\,(Kbits) the soft-mode processes all the tasks at the cloud, because in such a case fronthaul transmission is not the bottleneck of the delay, but when $B_k$ further increases, the transmission delay becomes prominent and parts of the  tasks should be locally processed by the fogs to avoid congestion in the fronthaul links.  Overall, from the above two examples, we see that the hard-mode tends to distribute more tasks at the fogs than the soft-mode.}

%\begin{table}
%	\centering
%	\caption{Ratio of tasks computed at the fog nodes vs. the task's size $B_{k}$.}
%	\label{table:task_at_fog_Bk}
%	\begin{tabular}{|c|c|c|c|c|c|c|}
%		\hline
%		$\substack{B_{k}\\ ({\rm Kbits})}$  &  20 & 60  & 100 &140 & 180 & 220 \\\hline
%		$\substack{{\rm Non-cooperative}\\(\%)}$ &  67 & 74 & 99 & 100 & 100 & 100\\\hline
%		$\substack{{\rm Cooperative}\\(\%)}$ &  0 & 5 & 15 & 22 & 30 & 35   
%		\\\hline
%	\end{tabular}\vspace*{1em}
%\end{table}

\section{Conclusion} \label{sec:conclude}
We have considered multiuser computation offloading in fog-radio access networks under both non-cooperative and cooperative offloading models. To guarantee the worst latency performance of all UEs, a joint communication and computation resource allocation problem is formulated 
as a min-max MINP. By leveraging the continuous reformulation, we have developed efficient inexact MM approach to the min-max problems. Simulation results have demonstrated that the  proposed offloading schemes are much better than some heuristic ones, and that the cooperative offloading is generally better than the non-cooperative one, owing to the cooperative gain from multiple fog nodes in the  fronthaul transmissions.

%show that our proposed method outperforms the minimum distance-based offloading strategy and the random offloading strategy.

%====HARD MODE========

%\begin{figure}[!h]
%	\centerline{\resizebox{.5\textwidth}{!}{\includegraphics{./figure/fog_comput_ratio_hard.pdf}}
%	}
%	\vspace{-5pt}
%	\caption{Ratio of tasks completed at fogs vs. $D_k$.} \label{fig:example5_hd}
%\end{figure}

%=====soft mode============
%\begin{figure}[!h]
%	\centerline{\resizebox{.4\textwidth}{!}{\includegraphics{./Fig_new/convergence_soft.pdf}}
%	}
%	\vspace{-5pt}
%	\caption{Convergence behavior of Algorithm~\ref{algorithm:2} with different initialization.} \label{fig:example1_soft}
%\end{figure}

%\begin{figure}[!h]
%	\centerline{\resizebox{.5\textwidth}{!}{\includegraphics{./figure/fog_comput_ratio_F_lmax_soft.pdf}}
%	}
%	\vspace{-5pt}
%	\caption{The ratio of tasks completed at the fogs vs. maximum computation capacity $F_{\ell,\max}$.} \label{fig:example3_soft}
%\end{figure}

%\begin{figure}[!h]
%	\centerline{\resizebox{.5\textwidth}{!}{\includegraphics{./figure/fog_comput_ratio_data_size_soft.pdf}}
%	}
%	\vspace{-5pt}
%	\caption{The ratio of tasks completed at the fogs vs. $B_k$.} \label{fig:example5_soft}
%\end{figure}

\section*{Appendix A. Proof of Theorem~\ref{theorem:1}}\label{appendix:cont_reformulation}
We first show that problem~\eqref{eq:main_reform} is a relaxation of \eqref{eq:main}. Since the objective of \eqref{eq:main} can be rewritten as
$ \alpha_{k,\ell}  (1-\beta_{k}) (\tau_{k,\ell}^T +   \tau_{k,\ell}^F)  +  \alpha_{k,\ell} \beta_{k} (\tau_{k,\ell}^T +  \tau_{k,\ell}^C )$, we set $\theta_{k,\ell}^F = \alpha_{k,\ell}  (1-\beta_{k})$ and $\theta_{k,\ell}^C = \alpha_{k,\ell} \beta_{k}$. It is easy to see that $\theta_{k,\ell}^F$ and $\theta_{k,\ell}^C$ are both nonnegative, and $\sum_{\ell=1}^L  \theta_{k,\ell}^F + \theta_{k,\ell}^C = \sum_{\ell=1}^L  (\alpha_{k,\ell}  (1-\beta_{k})+ \alpha_{k,\ell} \beta_{k}) = \sum_{\ell=1}^L\alpha_{k,\ell} = 1, \forall ~k $. Hence, the optimal solution of \eqref{eq:main} is a feasible solution of \eqref{eq:main_reform}.  Next, we show  that problem~\eqref{eq:main_reform} has an optimal solution, which is also a feasible solution of \eqref{eq:main}, thereby establishing equivalence of the two problems. Suppose that $(\tilde{\theta}_{k,\ell}^F, \tilde{\theta}_{k,\ell}^C)$ is an optimal solution of \eqref{eq:main_reform} and $\tilde{\tau}_{k,\ell}^X, ~X\in \{T, F, C\}, \forall~k,\ell$ is the corresponding latency calculated at the optimal solution. Without loss of generality, we assume $\tilde{\tau}_{k,\hat{\ell}}^T+\tilde{\tau}_{k,\hat{\ell}}^F \leq \tilde{\tau}_{k, {\ell}}^T+\tilde{\tau}_{k, {\ell}}^F$ and $\tilde{\tau}_{k,\hat{\ell}}^T+\tilde{\tau}_{k,\hat{\ell}}^F \leq \tilde{\tau}_{k, {\ell}}^T+\tilde{\tau}_{k, {\ell}}^C$ for all $\ell \neq \hat{\ell}$. In view of~\eqref{eq:main_reform_c} and \eqref{eq:main_reform_d}, it holds that
\begin{equation}\label{eq:opt_LB}
\sum_{\ell=1}^L  \left(   \tilde{\theta}_{k,\ell}^F  ( \tilde{\tau}_{k,\ell}^T+ \tilde{\tau}_{k,\ell}^F) +    \tilde{\theta}_{k,\ell}^C ( \tilde{\tau}_{k,\ell}^T+ \tilde{\tau}_{k,\ell}^C ) \right)  \geq \tilde{\tau}_{k,\hat{\ell}}^T+\tilde{\tau}_{k,\hat{\ell}}^F,
\end{equation}
That is, the choice of $\theta_{k,\hat{\ell}}^F = 1$ and $\theta_{k, {\ell}}^F = \theta_{k, {\ell}}^C = 0$ for all $\ell \neq \hat{\ell}$ is also optimal for \eqref{eq:main_reform}. In addition, since the lower bound in \eqref{eq:opt_LB} is independent of $\tilde{\tau}_{k, {\ell}}^F, \forall~\ell \neq \hat{\ell}$ and $\tilde{\tau}_{k, {\ell}}^C, \forall~\ell$, we can always set the communication and computational resources $f_{k,\ell}^F$, $C_{k,\ell}$ and $f_k^C$ appearing in $\tilde{\tau}_{k, {\ell}}^F, \forall~\ell \neq \hat{\ell}$ and $\tilde{\tau}_{k, {\ell}}^C, \forall~\ell$ to zero\,\footnote{Herein, we have by default assumed $\frac{0}{0}=0$.} without changing the optimal value $\tilde{\tau}_{k,\hat{\ell}}^T+\tilde{\tau}_{k,\hat{\ell}}^F$ of \eqref{eq:main_reform}. It is easy to verify that this particularly constructed optimal solution is also feasible, and attains the same objective value $\tilde{\tau}_{k,\hat{\ell}}^T+\tilde{\tau}_{k,\hat{\ell}}^F$ for problem~\eqref{eq:main}, if we  set $\alpha_{k,\hat{\ell}} = 1$, $\alpha_{k, {\ell}} =0, \forall~\ell \neq \hat{\ell}$ and $\beta_k =0$ in \eqref{eq:main_reform}. This completes the proof.  %\hfill $\blacksquare$

\section*{Appendix B. Proof of Theorem~\ref{theorem:stationary}} \label{appendix:proof_convergence}
Let us define
%\begin{align*}
%\bm x & \triangleq \{  \bm v_k, f_{k}^C, \bm \tau_{k}, \bm \theta_{k}   R_{k,\ell}, f_{k,\ell}^F, C_{k,\ell}\}_{k,\ell}, \\
%{\bm y} & \triangleq  \{ \bm u_{k,\ell}, w_{k,\ell}\}_{k,\ell}, \quad
%\nu(\bm x)  \triangleq   \max_{k\in \cal K} \Gamma_k(\bm x), \quad 
%\Gamma_k(\bm x)  \triangleq \bm \theta_k^T \bm \tau_k, \\
%\end{align*}
%\begin{align*}
%\tilde{\nu}(\bm x, \bm y; \bm x^{(t)}) & \triangleq \max_{k\in {\cal K}} \tilde{\Gamma}_k(\bm x,\bm y; \bm x^{(t)}), \\ \tilde{\Gamma}_k (\bm x,\bm y; \bm x^{(t)})  & \triangleq \frac{\| \bm \theta_k +\bm \tau_k \|^2}{2} - \left( \frac{1}{2} \| \bm \theta_k^{(t)} \|^2 + \frac{1}{2} \| \bm \tau_k^{(t)} \|^2 \right. \\
%&\left. +  (\bm \theta_k^{(t)})^T (\bm \theta_k - \bm \theta_k^{(t)}) +  (\bm \tau_k^{(t)})^T (\bm \tau_k - \bm \tau_k^{(t)})   \right).
%\end{align*}
\begin{align*}
& \bm x  \triangleq \{  \bm v_k, f_{k}^C, \bm \tau_{k}, \bm \theta_{k}   R_{k,\ell}, f_{k,\ell}^F, C_{k,\ell}\}_{k,\ell}, \\
&{\bm y}  \triangleq  \{ \bm u_{k,\ell}, w_{k,\ell}\}_{k,\ell}, \quad
\nu(\bm x)  \triangleq   \max_{k\in \cal K} \Gamma_k(\bm x), \quad 
\Gamma_k(\bm x)  \triangleq \bm \theta_k^T \bm \tau_k, \\
& \tilde{\nu}(\bm x, \bm y; \bm x^{(t)})  \triangleq \max_{k\in {\cal K}} \tilde{\Gamma}_k(\bm x,\bm y; \bm x^{(t)}), \\ 
&\tilde{\Gamma}_k (\bm x,\bm y; \bm x^{(t)})   \triangleq \frac{\| \bm \theta_k +\bm \tau_k \|^2}{2} - \left( \frac{1}{2} \| \bm \theta_k^{(t)} \|^2 + \frac{1}{2} \| \bm \tau_k^{(t)} \|^2 \right. \\
&\left. +  (\bm \theta_k^{(t)})^T (\bm \theta_k - \bm \theta_k^{(t)}) +  (\bm \tau_k^{(t)})^T (\bm \tau_k - \bm \tau_k^{(t)})   \right).
\end{align*}
Then, problem~\eqref{eq:dc_eqv} can be concisely expressed as
\begin{equation}\label{eq:proof_dc_sub}
\begin{aligned}
\min_{\bm x, \bm y} &~  \tilde{\nu}(\bm x, \bm y; \bm x^{(t)}) \\
{\rm s.t.} &~\zeta_{k,\ell}(\bm x, \bm y) \leq 0, \forall~k, \ell, \\
& ~ \psi_i(\bm x) \leq 0, ~i=1,\ldots, I,
\end{aligned}
\end{equation} 
where $\zeta_{k,\ell}(\bm x,\bm y) \triangleq R_{k,\ell}  - f_{k,\ell}(\bm u_{k,\ell}, w_{k,\ell}, \bm V )$ and $\psi_i(\bm x) \leq 0$ denotes the  constraints in \eqref{eq:main_reform2_b}-\eqref{eq:main_reform2_c} with $I$ being the total number of constraints. Without loss of generality, we assume $J^{(t)} = J$ for all $t$ in the following proof. With the above definitions and according to Algorithm~\ref{algorithm:1}, we have
\begin{subequations}\label{eq:proof_ineq}
	\begin{align}
	\nu(\bm x^{(t)}) &=  \tilde{\nu}(\bm x^{(t)}, \bm y^{(t)}; \bm x^{(t)})  \label{eq:proof_ineq_a}\\
	& = \tilde{\nu}(\bm x^{(t_0)}, \bm y^{(t_0)}; \bm x^{(t)})  \label{eq:proof_ineq_b}\\
	& \geq \tilde{\nu}(\bm x^{(t_1)}, \bm y^{(t_1)}; \bm x^{(t)}) \label{eq:proof_ineq_c}\\
	& \vdots \notag\\
	& \geq \tilde{\nu}(\bm x^{(t_J)}, \bm y^{(t_J)}; \bm x^{(t)}) \label{eq:proof_ineq_d} \\
	&\geq \nu(\bm x^{(t_J)})  \label{eq:proof_ineq_f}\\
	& = \nu(\bm x^{(t+1)}) \label{eq:proof_ineq_g}
	\end{align}
\end{subequations}
%\begin{subequations}\label{eq:proof_ineq}
%	\begin{align}
%	\nu(\bm x^{(t)}) &=  \tilde{\nu}(\bm x^{(t)}; \bm x^{(t)})  \label{eq:proof_ineq_a}\\
%	& = \tilde{\nu}(\bm x^{(t_0)}; \bm x^{(t)})  \label{eq:proof_ineq_b}\\
%	& \geq \tilde{\nu}(\bm x^{(t_1)}; \bm x^{(t)}) \label{eq:proof_ineq_c}\\
%	& \vdots \notag\\
%	& \geq \tilde{\nu}(\bm x^{(t_J)}; \bm x^{(t)}) \label{eq:proof_ineq_d} \\
%	&\geq \nu(\bm x^{(t_J)})  \label{eq:proof_ineq_f}\\
%	& = \nu(\bm x^{(t+1)}) \label{eq:proof_ineq_g}
%	\end{align}
%\end{subequations}
where \eqref{eq:proof_ineq_a} follows from the definitions of $\nu$ and $\tilde{\nu}$; \eqref{eq:proof_ineq_b} is because $\bm x^{(t)}$ is chosen as initialization of $\bm x^{(t_0)}$; \eqref{eq:proof_ineq_c} follows from the descent property of block-coordinate minimization; \eqref{eq:proof_ineq_f} is because $\tilde{\nu}$ majorizes $\nu(\bm x)$; \eqref{eq:proof_ineq_g} follows from the definition of $\bm x^{t+1}$ in Algorithm~\ref{algorithm:1}. 
Therefore, the iterates $\{\bm x^{(t)}\}_t$ generated by Algorithm~\ref{algorithm:1} yield a non-increasing objective values for problem~\eqref{eq:main_reform2}. Since problem~\eqref{eq:main_reform2} is lower bounded below, by monotone convergence theorem,   $\nu(\bm x^{(t)})$ must converge to some finite value, i.e.,
\[ \lim_{t\rightarrow \infty} \nu(\bm x^{(t)}) = \nu^\star >-\infty. \]
From \eqref{eq:proof_ineq_a}-\eqref{eq:proof_ineq_g}, we also have
\begin{equation} \label{eq:key_0}
\lim_{t\rightarrow \infty} \tilde{\nu}(\bm x^{(t)},\bm y^{(t)}; \bm x^{(t)} ) = \nu^\star.
\end{equation}
Consider a converging subsequence $(\bm x^{(t_{j})}, \bm y^{(t_{j})})_j$ of $(\bm x^{(t)}, \bm y^{(t)})_t$ such that
\[  \lim_{j\rightarrow \infty} (\bm x^{(t_{j})}, \bm y^{(t_{j})})  = (\bar{\bm x}, \bar{\bm y}).  \]
Now, by taking limit along the converging subsequence $({\bm x}^{(t_{j})}, {\bm y}^{(t_{j})})_j$ on both sides of \eqref{eq:key_0}, we get
\begin{equation}\label{eq:key_1}
\tilde{\nu}(\bar{\bm x}, \bar{\bm y}; \bar{\bm x} ) = \nu^\star.
\end{equation}
From the descent property in~\eqref{eq:proof_ineq}, we also have
%\begin{equation}\label{eq:key_2}
%\begin{aligned}
%\tilde{\nu}({\bm x},  {\bm y}^{(t_{j})}; {\bm x}^{(t_j)} ) & \geq \tilde{\nu}({\bm x}^{(t_{j_1})},  {\bm y}^{(t_{j})}; {\bm x}^{(t_j)} ) \\
%& \geq  \tilde{\nu}({\bm x}^{(t_{j_1})},  {\bm y}^{(t_{j_1})}; {\bm x}^{(t_j)} ) \\
%& \geq \ldots \geq \nu^\star
%\end{aligned}
%\end{equation} 
\begin{equation}\label{eq:key_2}
\tilde{\nu}({\bm x},  {\bm y}^{(t_{j})}; {\bm x}^{(t_j)} )  \geq   \nu^\star, ~~\forall~ \bm x \in {\cal F}(\bm y^{(t_j)}),
\end{equation} 
where ${\cal F}(\bm y^{(t_j)})$ denotes the feasible set of problem~\eqref{eq:proof_dc_sub} when fixing $\bm y = \bm y^{(t_j)}$.
By taking limit along the converging subsequence $({\bm x}^{(t_{j})}, {\bm y}^{(t_{j})})$ on both sides of \eqref{eq:key_2}, we get
\begin{equation}\label{eq:key_3}
\tilde{\nu}({\bm x},  \bar{\bm y}; \bar{\bm x}) \geq \nu^\star, \quad \forall~{\bm x} \in {\cal F}(\bar{\bm y})
\end{equation}
Combining \eqref{eq:key_1} and \eqref{eq:key_3}, we obtain the following key inequality:
\begin{equation}\label{eq:key_4}
\tilde{\nu}({\bm x},  \bar{\bm y}; \bar{\bm x})  \geq \tilde{\nu}(\bar{\bm x},  \bar{\bm y}; \bar{\bm x}) , \quad \forall~{\bm x} \in {\cal F}(\bar{\bm y}).
\end{equation}

On the other hand, since ${\bm y}^{(t_{0})}$ is obtained by minimizing problem~\eqref{eq:proof_dc_sub} with fixed ${\bm x}^{(t_{0})}$, we have
\begin{equation}\label{eq:key_5}
\tilde{\nu}({\bm x}^{(t_0)},{\bm y}^{(t_{0})}; {\bm x}^{(t)} ) \leq \tilde{\nu}({\bm x}^{(t_0)},{\bm y}; {\bm x}^{(t)} ), ~ \forall~\bm y 
\end{equation} 
Again, by taking limit along the converging subsequence $(\bm x^{(t_{j})}, {\bm y}^{(t_{j})})_j$ on both sides of \eqref{eq:key_5}, we get another key inequality
\begin{equation}\label{eq:key_6}
\tilde{\nu}(\bar{\bm x},{\bm y}; \bar{\bm x} ) \geq \tilde{\nu}(\bar{\bm x}, \bar{\bm y}; \bar{\bm x} ) , \quad \forall~\bm y.
\end{equation}

Next, we will complete the proof by exploiting the two key inequalities in~\eqref{eq:key_4} and \eqref{eq:key_6}. Specifically, the inequality~\eqref{eq:key_4} implies that $\bar{\bm x}$ is an optimal solution for the following problem:
\begin{equation}\label{eq:proof_x_opt}
\begin{aligned}
\min_{\bm x} &~  \tilde{\nu}(\bm x, \bar{\bm y};\bar{\bm x}) \\
{\rm s.t.} &~\zeta_{k,\ell}(\bm x, \bar{\bm y}) \leq 0, \forall~k, \ell, \\
& ~ \psi_i(\bm x) \leq 0, ~i=1,\ldots, I.
\end{aligned}
\end{equation}
Hence, $\bar{\bm x}$ must satisfy the KKT conditions of problem~\eqref{eq:proof_x_opt}, which are listed below.
\begin{equation}\label{eq:kkt_check}
\begin{aligned}
& \bm 0  \in \partial_{\bm x} \tilde{\nu}(\bar{\bm x}, \bar{\bm y}; \bar{\bm x}) + \textstyle \sum_{k,\ell} \kappa_{k,\ell} \nabla_{\bm x}\zeta_{k,\ell}(\bar{\bm x}, \bar{\bm y}) + \textstyle \sum_{i} \eta_i \nabla_{\bm x}\psi_i(\bar{\bm x})  \\
& \kappa_{k,\ell} \zeta_{k,\ell}(\bar{\bm x}, \bar{\bm y}) = 0,  \\
&  \eta_i \psi_i(\bar{\bm x}) = 0, \\
& \zeta_{k,\ell}(\bar{\bm x}, \bar{\bm y}) \leq 0, ~\forall~k,\ell,\\
& \psi_i(\bar{\bm x})  \leq 0, ~\forall~i,\\
& \kappa_{k,\ell} \geq 0, \quad \eta_i \geq 0, ~\forall~k,\ell,i,
\end{aligned}
\end{equation} 
where $\kappa_{k,\ell}$ and $\eta_i$ are Lagrangian multipliers; $\partial_{\bm x} \tilde{\nu}$ denotes the subdifferential of $\tilde{\nu}$. Moreover, the inequality~\eqref{eq:key_6} implies that  $\bar{\bm y}$ is an optimal  solution of problem~\eqref{eq:proof_dc_sub} for fixed $\bar{\bm x}$. Recall that for fixed $\bm x$, the optimal $\bm y$ can be {\it uniquely} computed in closed form by~\eqref{eq:wmmse_update}. Therefore, the optimal $\bar{\bm y}$ of problem~\eqref{eq:proof_dc_sub} takes the form of~\eqref{eq:wmmse_update} (with $\bm V$ replaced by $\bar{\bm V}$). Now, by substituting this specific $\bar{\bm y}$ into $\zeta_{k,\ell}(\bar{\bm x}, \bar{\bm y})$, one can easily verify that the following holds:
\begin{equation}\label{eq:kkt_check_key}
\begin{aligned}
\zeta_{k,\ell}(\bar{\bm x}, \bar{\bm y})  & = R_{k,\ell} - \phi_{k,\ell}(\bar{\bm x})   \\
\nabla_{\bm x}\zeta_{k,\ell}(\bar{\bm x}, \bar{\bm y})  & =   \nabla_{\bm x} \phi_{k,\ell}(\bar{\bm x}),
\end{aligned}
\end{equation}
where $ \phi_{k,\ell}$ is defined in~\eqref{eq:main_reform2}. Notice that we have used Danskin's theorem~\cite{Bertsekas} to obtain the second equation in \eqref{eq:kkt_check_key}. By substituting~\eqref{eq:kkt_check_key} into \eqref{eq:kkt_check}, we almost obtain the KKT conditions of problem~\eqref{eq:main_reform2}, except for one remaining issue to verify, i.e., $\partial_{\bm x} \tilde{\nu}(\bar{\bm x}, \bar{\bm y}; \bar{\bm x}) = \partial_{\bm x}  {\nu}(\bar{\bm x})$. This can be  shown as follows.   Notice that 
\begin{equation}\label{eq:partial_gradient}
\begin{aligned}
\partial_{\bm x} \tilde{\nu}(\bar{\bm x}, \bar{\bm y}; \bar{\bm x}) = & {\rm Conv }\{\cup_{k \in \cal A}  \nabla_{\bm x} \tilde{\Gamma}_k(\bar{\bm x}, \bar{\bm y}; \bar{\bm x})\} \\
= & {\rm Conv }\{\cup_{k \in \cal A}  \nabla_{\bm x} \Gamma_k(\bar{\bm x})\}\\
= & \partial_{\bm x}  {\nu}(\bar{\bm x})
\end{aligned}
\end{equation} 
where ${\rm Conv}\{ \cdot\}$ denotes the convex hull, and ${\cal A}$ represents the set of active indices satisfying  $\tilde{\Gamma}_k(\bar{\bm x}, \bar{\bm y}; \bar{\bm x}) = \tilde{\nu}(\bar{\bm x}, \bar{\bm y}; \bar{\bm x})$. The second equality in \eqref{eq:partial_gradient} is due to the fact that $\tilde{\Gamma}_k( {\bm x}, \bar{\bm y}; \bar{\bm x})$ is the tight approximation of $\Gamma_k( {\bm x})$ up to first order at the point $\bm x= \bar{\bm x}$. This completes the proof.

\bibliographystyle{IEEEtran}
%\footnotesize
\bibliography{ref2}
%\vspace*{-8pt}
%\input{fix.bbl}

\end{document}